\documentclass[unsortedaddress,pre]{revtex4-2}
\usepackage{graphicx}
\usepackage{epstopdf}
\usepackage{float,subfigure}
\usepackage{amsmath,amssymb}
\usepackage{natbib}
\usepackage{xcolor}

\usepackage{dcolumn}%
\usepackage{bm}%

\def\be{\begin{equation}}
\def\ee{\end{equation}}
\DeclareMathOperator*{\argminB}{argmin}
\DeclareRobustCommand{\vect}[1]{
  \ifcat#1\relax
    \boldsymbol{#1}
  \else
    \mathbf{#1}
  \fi}

\newcommand{\secref}[1]{\hyperref[sec:#1]{Section #1}}

\newcommand\F{{\mathcal F}}

\newcommand\N{{\mathcal N}}
\newcommand\I{{\mathbb I}}
\newcommand\IC{{\mathcal I}}
\newcommand\n{^{\scriptscriptstyle (N)}}

\newcommand\x{{\bf x}}

\newcommand\e{{\bf e}}
\newcommand\f{{\bf f}}
\newcommand\Tr{\text{Tr}}

\newcommand\RR{\mathcal{R}}
\newcommand\R{\mathbb{R}}
\newcommand\egaldef{\stackrel{\mbox{\upshape\tiny def}}{=}}

\draft 

\begin{document}


\title{Building causation links in stochastic nonlinear systems from data} 



\author{Sergio Chibbaro}
\affiliation{Universit\'e Paris-Saclay, CNRS, UMR 9015, LISN, 91190 Gif-sur-Yvette, France}
\author{Cyril Furtlehner}
\affiliation{INRIA-Saclay,Universit\'e Paris-Saclay, LISN, 91190 Gif-sur-Yvette, France}

\author{Th\'eo Marchetta}
\affiliation{INRIA-Saclay,Universit\'e Paris-Saclay, LISN, 91190 Gif-sur-Yvette, France}

\author{Andrei-Tiberiu Pantea}
\affiliation{Universit\'e Paris-Saclay, CNRS, UMR 9015, LISN, 91190 Gif-sur-Yvette, France}

\author{Davide Rossetti}
\affiliation{Politecnico di Torino, Corso Duca degli Abruzzi 24, I - 10129, Torino, Italy}
\affiliation{Universit\'e Paris-Saclay, UMR 9015, LISN, 91190 Gif-sur-Yvette, France}



\begin{abstract}
  Causal relationships play a fundamental role in understanding the world around us. The ability to identify and understand cause-effect relationships is critical to making informed decisions,
  predicting outcomes, and developing effective strategies.
 However, deciphering causal relationships from observational data is a difficult task, as correlations alone may not provide definitive evidence of causality. 
 In recent years, the field of machine learning (ML) has emerged as a powerful tool, offering new opportunities for uncovering hidden causal mechanisms and better understanding complex systems. 
 In this work, we address the issue of detecting the intrinsic causal links of a large class of complex systems in the framework of the response theory in physics. We develop some theoretical
 ideas put forward by~\cite{aurell2016causal}, and technically we use state-of-the-art ML techniques to build up models from data.
 We consider both linear stochastic and non-linear systems. Finally, we compute the asymptotic efficiency of the linear response based causal predictor in a case of large scale
 Markov process network of linear interactions.
\end{abstract}


\maketitle

%

\section{Introduction}

The general definition of causality may be slippery and has been a subject of scholar study since ever~\cite{aristotle1984complete,hume1896treatise,russell1912notion}.
However, the recognition and clarification of some kind of cause-effect relationship between variables, events or objects are fundamental questions of natural and social
sciences \cite{hlavavckova2007causality,pearl2009causality,zhang2011causality}. 
The motivation to analyze such a problem is very high, since in most practical cases we need to quantify the strength of a possible causal
relation in order to explain past events, control present situations and predict future
consequences~\cite{marconi2008fluctuation,zeng2011network,friedrich2011approaching,baldovin2018role,ferretti2020building}.
One clear instance of the question is the interplay between carbon emissions and temperature, or to understand the possible relation between the contact with some
substance and the subsequent development of a disease. 
Causal analysis is a distinguished branch of statistics~\cite{yedidia2003understanding,pearl2009causality}, with a possible impact on machine learning~\cite{wang2022respecting}.

The crucial point is to define properly the causality from a physical standpoint.
In classical physics this idea is often associated with a principle of legal
determinism, in which an antecedent uniquely determine the following state of the system, an idea rooted in classical mechanics~\cite{kojeve1990idee,cohen2000collapse}. 
However, in this framework the notion of causality leads to considerable epistemological difficulties, and the notions of cause and effect are in fact not invoked.
Instead, in a recent work~\cite{aurell2016causal} it has been shown that there is a conceptual parallelism between 
the statistical causal analysis and long-time response in physics~\cite{marconi2008fluctuation}.
There is still a crucial difference. In physics, the intervention on a system may cause some response only after some time, whereas statistical
analysis of causality assumes immediate effects, the notion of time ordering being absent.
In this sense, statistical analysis should be easier to work but generally not of physical relevance.
The possibility to use the response theory to establish causal links is particularly interesting for those cases for which laboratory experiments are not possible, for instance in geophysical problems.
As a matter of fact, in many realistic phenomena one is confronted to a time-series of few observables~\cite{kantz2003nonlinear,assaad2022survey}.

On a general basis, we may formulate the problem 
as follows:  a time series $\{x_{1t},x_{2t},\ldots,x_{dt}\}$ of $d$ variables represent an observable system $\x_t$, 
and one wishes to determine unambiguously whether the behavior of $x_i$ has been 
influenced by $x_j$ during the dynamics, without knowing the underlying 
evolution laws. 
A first natural way to look at this problem is to look at correlations between variables $C_{ij}(t)=\langle x_{it} x_{j0}\rangle$, since a causal link is expected to manifest as a non-zero value 
for it for some $t>0$, at least following  intuition. 
Yet, it is well known that correlation 
does not imply causation, for instance,  both  $x_i$ 
and $x_j$ may be influenced by common-causal 
variables~\cite{pearl2009causality,simon1954spurious}.

More reliable tools to point out causal effects between two 
variables are the Granger causality (GC) test~\cite{granger1969investigating,seth2007granger}, and  transfer entropy (TE) approaches.
GC allows to determine whether the knowledge of the past history of 
$x_j$  increases the ability to predict future values of $x_i$~\cite{cadotte2008causal,barrett2010multivariate}. 
TE is based on the definition of 
a degree of information exchange from $x_j$ to $x_i$, which quantifies 
the loss of information about $x_i$ that one experiences if $\{x_{jt}\}$ 
is ignored~\cite{schreiber2000measuring,bossomaier2016transfer}.
However, from a physical point of view, these definitions may be not quite satisfactory.
In physics, two variables are usually considered to have a mutual cause-effect relationship if an 
external action on one of them results in a change of the observed value of the 
second~\cite{barnett2009granger,aurell2016causal}.
As explained shortly, GE and TE provide a weaker information, whether, and to what extent, the knowledge of a certain variable is 
useful to the actual determination of future values of another. 
This kind of causality is called ``observational'', while
the stronger physics-oriented one is termed ``interventional''.
In a different framework, the latter is also favored in computer science by Pearl~\cite{pearl2009causality}. 

In some recent works~\cite{baldovin2020,baldovin2022extracting},
the link between the interventional causal link and the physical response has been investigated, considering also applications of the utmost importance.
However, some issues may hinder the possibility of using such approach, both from a theoretical and a practical point of view.
This approach appears to be perfectly suited for linear Markov systems.
It may fail in the case of nonlinear and/or non-Markovian systems, which however represent most of the relevant biological and physical ones.
In particular,  computing directly the response of a system implies the possibility to apply an infinitesimal perturbation to the system and observe the possible following statistical drift. The experiment must be repeated many times to have statistical convergence.
Yet, in actual problems either we have a system we cannot act upon, either we have just a time-series and such experiments cannot be performed.
Solely in the case of linear systems, covariance matrices are enough to reconstruct the entire response.
Furthermore, from a more practical point of view, the data available are not necessarily enough to have a robust assessment of the response function, even in the linear case.

In this work, we address the issue of detecting causality in general nonlinear Markov or chaotic systems, from the physical perspective.
The goal is to leverage the capability of machine learning algorithms to build up models from data.
Machine learning (ML) has been notably developed in the field of computer vision~\cite{goodfellow2016deep}, and it has brought new perspectives in the data assimilation and analysis of complex physical systems~\cite{brunton2022data,brunton2020machine,buzzicotti2021reconstruction,brajard2020combining}. 

To the best of our knowledge, we present here the first study of the physical response problem in terms of a machine learning approach. 
In particular, we leverage this framework to establish causal links among dynamical variables.
In this way, we go beyond previous studies and we generalize to very large Markov system and to a full nonlinear dynamics.
We investigate whether it is possible to infer causal links purely from data,
when these data are generated by complex noisy/chaotic systems. We consider both the case in which no prior structure of the system is given,
and when some physical information is used to design the data-driven approach.
More specifically,  we leverage  ML techniques to estimate the response function in various regimes, i.e. linear and non-linear with or without cofounders;
we set the basis to analyze the large scale behavior, by providing explicit formulas for a generic linear Markov process in the asymptotic limit, using the Random Matrix formalism.
Since neural networks are very expressive nonlinear models, we are able to go beyond the linear response approximation considered in previous studies and to get at least
qualitative estimates of the causal links, even in the general fully nonlinear cases.


The paper is organized as follows:
in Section.~\ref{sec II} we discuss the two main different types of approaches to  causal inference, namely observational and interventional, 
  and how the latter can be virtually done via generalized response theory. In Section.~\ref{sec:resp_theory} we recall the basics of generalized response theory, both
  for linear and non-linear settings. In Section.~\ref{sec:Markov}
  we revisit  the linear and non-linear stochastic examples studied  in~\cite{baldovin2020} with ML approaches.
  In Section.~\ref{sec:chaotic_systems} we consider a setting with chaos by tackling the Lorenz model also with ML techniques.
Finally in Section.~\ref{sec:theory} we analyze theoretically the efficiency of causal indicator based on linear response theory
for large scale causal graphs with help of random matrix theory(RMT).

\section{Definition of Causality}\label{sec II}

As anticipated in the introduction, a consistent and relevant definition of a causality relation may be slippery, and it is crucial to define properly the terms of the problem.
To obtain information about causal relations there are basically two broad types of approaches: the interventional and the observational one.
Both provide some information about the presence or absence of cause and effect, but they differ in how this information is expressed.
We quickly explain both approaches in the language of statistical physics, following the same line depicted in a recent work~\cite{baldovin2020}.


For the observational setting, two widely used methods to extract information are the Granger
causality (GC)~\cite{wiener1956nonlinear,granger1969investigating,Seth:2007} for linear causal relations and the transfer entropy (TE)~\cite{schreiber2000measuring,kantz2003nonlinear},
which generalizes the former to non-linear situations.

GC can be quantified and measured computationally, usually in the context of linear regressive models. It makes two statements about causality: 1) the cause occurs before the effect and 2) the cause contains information about the effect that is unique and does not occur in any other variable. Consequently, the causal variable can help to anticipate the effect variable after other data have been used first \cite{lindner2019comparative}.
This is a regression-based interpretation of past data that compares the statistical uncertainties of two predictions.
For example, let us consider the time series of two events \(x_j(t)\) and \(x_i(t)\), we can use a model where only the past
history of \(x_j\) is included to predict future values of itself: 
\begin{equation}
x_j(t) = \sum_{k=1}^T B_{j,k}x_j(t-k)+\epsilon_j(t)
\end{equation}
where \(B\) is the auto-regressive coefficient and \(\epsilon_j\) is the prediction error.
A second prediction may be obtained using a  model, in which the past of \(x_i\) is also included in the model for predicting future values of \(x_j\):
\begin{equation}
x_j(t) = \sum_{k=1}^T [A_{ji,k}x_i(t-k)+A_{jj,k}x_j(t-k)+\epsilon_{j|i}(t)]
\end{equation}
where, as before, the \(A\)'s are the auto-regressive coefficients and \(\epsilon_{j|i}\) is the prediction error on \(x_j\) associated with the knowledge of variable \(x_i\).
If we register an improvement in prediction, i.e. a reduction in the variance of prediction errors, we say that \(x_i\) is a Granger cause of \(x_j\). G-Causality may be quantified quantified as:
\begin{equation}
F_{x_i \xrightarrow[]{}x_j } = \log{\frac{Var(\epsilon_j)}{Var(\epsilon_{j|i})}}
\end{equation}
So if this value is positive, we have an improvement in predictive accuracy. GC can also be seen as a statistical hypothesis test.

TE derives causality from an information theory interpretation. It is a specific version of a Kullback-Leibler entropy for conditional probabilities.
It is a non-parametric statistic that measures the amount of information exchanged between  \(x\) and \(y\). More specifically, it indicates the amount of uncertainty
reduced in future values of \(y\) given knowledge of both the past values of \(x\) (and given past values of \(y\)) with respect to the knowledge of
the past $y$ alone~\cite{kantz2003nonlinear,hlavavckova2007causality}. 

As affirmed by Granger himself, it is generally agreed that G-causality does not capture all aspects of causality, but enough to be worth considering in an empirical test~\cite{Seth:2007}.
The same statement hold for transfer entropy~\cite{james2016information}.
A key observation regarding all these observational methods is their complete reliance on selecting the right variables. Causal factors excluded from the regression model cannot be
reflected in the results. Consequently, Granger causality or Transfer Entropy should not be interpreted as direct representations of physical causal mechanisms~\cite{Seth:2007}.
In this sense, this type of causation is not necessarily satisfactory from a physical point of view and for this reason, is discarded in some situations in favor of a more ``interventional" approach.

Interventional causality is more easily linked to the physical cause-effect relationship. The state of a variable is changed to manipulate the system and see if there is a reaction.
Since we are in a statistical framework, it states that given a time series characterized by the vector \(\boldsymbol{x}_t\) consisting of \(n\) entries, a perturbation of the variable \(x_j\) at
time \(0\), \(x_{j0} \xrightarrow{} x_{j0} + \delta x_{j0}\), on average, causes a change in another variable \(x_{kt}\) with \(t > 0\). 
In causal statistics~\cite{pearl2009causality,judea2010introduction},  interventional causality 
relies on causal Bayesian networks and directed acyclic graphs (DAG), and aims to understand the representation of the causal structure and the response to external or spontaneous
changes in variables. The main technical tool is the probability conditioned on an intervention, that is a change of some variable. 
This formalism permits to determine for a known graph whether one has sufficient data to correctly estimate the conditional probability, and in this case indicate what the conditional
probability model would be for a modified model. The Pearl's approach may therefore be used to assess if enough data are available to answer a formalized question.

In~\cite{aurell2016causal}, it was suggested that conceptually the Bayesian belief network used in causal statistics may be replaced in physical problems by a dynamical probabilistic model,
which should contain the same dependencies.
In this framework, the probability conditioned on a given intervention is related to the long-time response of the system to it.
As a consequence, there is a link between the causality intended as a time-ordered relationship where a change in a variable determines a change on another variable at successive times.
This idea can be mathematically formalized saying that there is a causal relation if, given a smooth function \(\F(x)\), the relationship holds:
\begin{equation}
\frac{\overline{\delta \F(x_{kt})}}{\delta x_{j0}} \neq 0
\label{eq:resp}
\end{equation}
which means that a perturbation of the variable \(x_{j0}\) at the time \(0\) leads to a non-zero average variation \(\F(x_{kt})\) (carried out over many
realizations of the experiment) with respect to its unperturbed evolution \cite{baldovin2020}.
It is possible to appreciate that the formulation given by Eq. (\ref{eq:resp}) of the causality naturally leads to the statistical response of the
system~\cite{kubo2012statistical,marconi2008fluctuation,livi2017nonequilibrium}.

In this work we focus on this notion of interventional causality in the statistical physics framework.

\section{Response theory Framework}\label{sec:resp_theory}

Starting from the general definition Eq. (\ref{eq:resp}), a connection can be made with the response theory,
when the system admits an invariant measure and considering that the variation \(\delta x_{j0}\) is small enough. 

We recall the main ideas and results of response theory.
We focus in particular on the choice $\F(x_{kt})=\delta x_{kt}$.
Complements can be found in appendix, while a more detailed exposition can be found in Ref.~\cite{marconi2008fluctuation}.
Consider a Markov process $\x_t=(x_{1t}, ..., x_{dt})$ of dimension $d$ whose invariant measure $\rho$
is smooth and nonvanishing. Given a (small) perturbation $\delta \x_0 = (\delta x_{10}, ..., \delta x_{d0})$ at time $t=0$, its effects at time $t$ is obtained by measuring
the difference between the vector $\x_t$ in the original dynamics and in the perturbed one, 
on average. Formally we compute
 \begin{equation}
 \overline{ \delta x_{kt}} = \langle x_{kt} \rangle_p -\langle x_{kt} \rangle\,,
 \end{equation} 
where $ \langle \cdot \rangle_p$ and $ \langle \cdot \rangle$ indicate the average over many realizations of the perturbed and of the original dynamics, 
respectively.

As shown in appendix \ref{app:resp-teo}, it is possible to derive a closed formula for the response of a dynamical system  valid under rather general 
hypotheses~\cite{falcioni1990correlation}; in particular, in its derivation no assumption of detailed 
balance is used, meaning that the formula also holds 
for out-of-equilibrium systems in stationary states.
The result is the following, for \(\boldsymbol{x}_t \)  a stationary process characterized by an
invariant probability density function \(\rho(\boldsymbol{x})\), we can write the following relation:
\begin{equation}
  \RR_{j\to k}(t) \equiv \lim_{\delta x_{j0} \to 0} \frac{\overline{\delta x_{kt}}}{\delta x_{j0}} = -
  \left\langle x_{kt} \frac{\partial \ln{\rho(\x)}}{\partial x_j} \left. \vphantom{\frac{a}{b}} \right|_{\x_0} \right\rangle
\label{eq:response}
\end{equation}
where \(\RR(t)\) represents the linear response matrix of the system under consideration at time \(t\), and the average is taken over the joint stationary probability distribution
function \(p(\x_0, \x_t)\). 

If we consider now the special case of a linear dynamics ruled by a discrete-time stochastic evolution
\begin{equation}
\label{eq:evo_app}
 \x_{t+1}=A \x_{t} + B \vect{\eta}_t
\end{equation}
where $A$ and $B$ are $d \times d$ matrices and $\vect{\eta}_t$ is a $t$-dependent 
vector of delta-correlated random variables with zero mean,
the relation  (\ref{eq:response}) becomes simply
\begin{equation}
\label{eq:resplin}
 \RR_{i\to j}(t)=[A^t]_{ji}= [C_tC_0^{-1}]_{ji}\,,
\end{equation} 
valid for linear Markov systems at discrete times, where we have introduced the covariance matrix $C_t=\langle \x_t \x^{T}_0\rangle$.
This is a key result of linear response theory~\cite{marconi2008fluctuation}. 
Therefore, for a linear system the knowledge of the covariance matrix fully characterizes the response of the system and thus the causality links.


In general, it is not possible to have access to the joint probability distribution function $p(\x_0, \x_t)$,
while the computation based on the covariance matrix is valid only for linear systems.
For these reasons, in this work, we compute the response using another method, linked to the general definition~(\ref{eq:response}).
This is more fundamental from a conceptual point of view~\cite{ciccotti1975direct,j2007statistical}, since we use
directly the  definition of transport coefficients: one perturbs the
system with an external force or imposes driving boundary
conditions and observes the relaxation process.  
In our numerical simulations, we compute $\RR_{i\to j}(t)$ by perturbing the variable $x_i$ at time $t=t_0$ with a small
perturbation of amplitude $\delta x_i(0)$ and then evaluating the 
separation $\delta x_j(t|t_0)$ between the two trajectories ${\bf
x}(t)$ and ${\bf x}'(t)$ which are integrated up to a prescribed time
$t_1=t_0+\Delta t$.  At time $t=t_1$ the variable $x_i$ of the
reference trajectory is again perturbed with the same $\delta x_i(0)$,
and a new sample $\delta {\bf x}(t|t_1)$ is computed and so forth.
The procedure is repeated $N \gg 1$ times and the mean response is
then evaluated:
\begin{equation} 
\label{eq:resp-num}
\RR_{i\to j}(\tau)= {1 \over N} \sum_{s=1}^N 
{ {\delta  x_j(t_s+\tau|t_s)} \over  {\delta  x_i(t_s)}}  \,\, .
\end{equation}
where it is assumed that the delay $\Delta t = t_{s+1}-t_s$ between two consecutive  examples is much larger than the auto-correlation time of the process.

\section{Low dimensional Markov systems}\label{sec:Markov}
Let us start with the $3$-dimensional linear process example considered in~\cite{baldovin2020}, where Eq.~(\ref{eq:evo_app}) is specified by
\[
A = 
\left(
\begin{matrix}
a & \epsilon & 0 \\  
a & a & 0 \\  
a & 0 & a
\end{matrix}
\right)
\]
and  $B = \sigma \I$ representing isotropic noise, with standard deviation $\sigma$.   
The parameters \(a\), \(\sigma\) and \(\epsilon\) are constant in time,   
with default valued \(a=0.5\) and \(\sigma=1\) in the following while the linear response is studied by varying $\epsilon$. 
Thanks to the linearity of the system the interventional causal relationships relies on the sole auto-correlation of the signal as seen in Eq.~(\ref{eq:resplin}).
In this simple example the interventional causality of the system can be computed exactly~\cite{baldovin2020} so that we can use this system
as a benchmark to assess different data-driven techniques. 

Here we focus on the possibility to find the same results directly from data, without prior knowledge of the model.  
We consider thus to  have the time-series of a system, without knowledge of the underlying model, though we make the assumption that the system is linear, and use these data to empirically estimate
the linear response~(\ref{eq:resplin}) which actually corresponds to solving a Linear Regression~\cite{bishop2006pattern}.
This can be done in two different ways:
either we estimate the Markov matrix $A$ (Markov regression) from the data and compute the linear response in~(\ref{eq:resplin}) by taking powers of this estimate;
either we directly compute the linear response~(\ref{eq:resplin}) (linear regression) at finite time by estimating the auto-correlation of the process.
Note that the second approach is more general as it applies to partially observed system while the former assumes a fully observed system.
In addition the linear response being equivalent to a linear regression, it can be regularized by adding a penalization on the regression coefficients $w_{ij}$
which actually coincides with the response coefficient $\RR_{j\to i}(t)$ for a given $t$.
Hence we can combine these two inference methods with the Lasso Regularization, also known as $L_1$ regularization, in order to filter out small spurious regression coefficients and select only the
relevant ones~\cite{mehta2019high,brunton2022data}. 
This consists in adding a term of the form
$     \text{L1 Penalization} = \lambda \sum_{i,j=1}^{n}\lvert w_{ij} \rvert$ to the loss function, with $\lambda$ the $L_1$ penalty. In our situation it should help to decide whether
a regression coefficient corresponding to $\RR_{y\to x}$ is zero or not. 
\begin{figure}[H]
    \centering
     \includegraphics[height=0.4\textwidth]{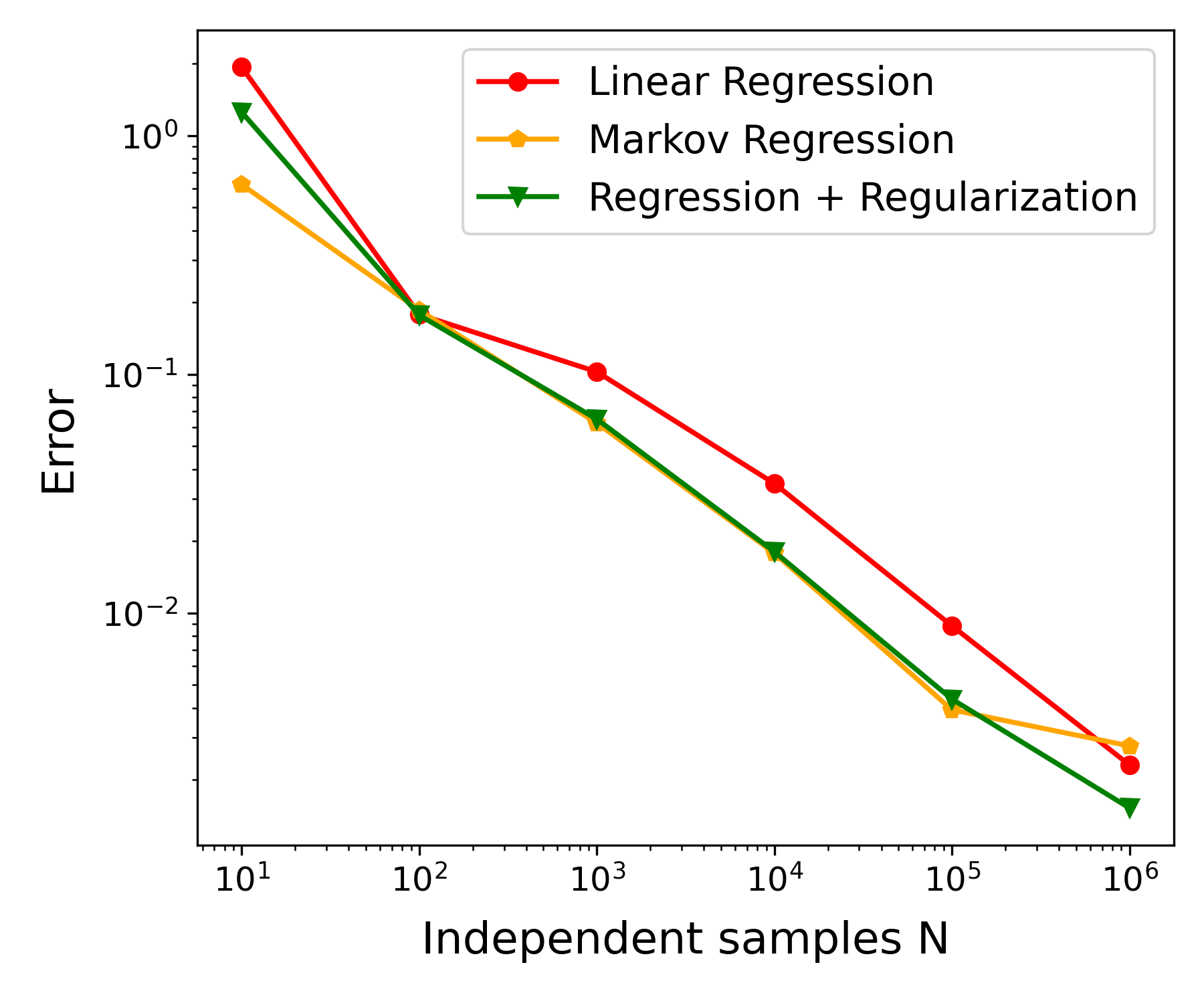}
     \caption{Plot of the square error of the reconstruction of the matrix \(A\) for  four different methods. The resulting error for each method was calculated for a different number
       of training samples.}
    \label{fig:2}
\end{figure}
We estimate the global error for the different methods as the following: we compute the square root of the sum of the quadratic single error over each coefficient is given in
Fig.\ref{fig:2}. All methods give similarly satisfactory  results and converge as expected by law of large numbers as the number of independent samples increases. The $L_1$
regularization seems to reduce to fluctuations. 
So if the system is not explicitly known different machine learning techniques may still be used to reconstruct the Markov process and the response function. 

Now, we turn to a non-linear generalization involving  three interacting particles in one dimension also studied in~\cite{baldovin2020}.
These particles, described by variables \(x\), \(y\) and \(z\), are under the influence of a quartic potential, and we assume a dynamics characterized by overdamping.
The evolution of the system  is determined by the following equations:
\begin{eqnarray}
     \dot{x} &=& -U'(x)-(x-y)+d\eta^{(x)}, \nonumber \\
     \dot{y} &=& -U'(y)-(y-x)-(y-z)+d\eta^{(y)}, \\ 
     \dot{z} &=& -U'(z)-(z-y)+d\eta^{(z)}.\nonumber
\label{eq:nl-mark}
\end{eqnarray}
with the potential $U(x)=(1-r)x^2+rx^4$.
The components of the noise  \(d\eta\) are iid centered normal variables with variance equal to the used time step.
The parameter \(r\) tune the level of non-linearity of the dynamics. When \(r=0\) we recover a linear dynamics.
For \(r\ge 1\)  the potential \(U\) takes on a pronounced double-well form.

Consistently with the study conducted in~\cite{baldovin2020}, we consider \(r\in\{1,2.4\}\) for our analysis. 
For simulations, we take a time step of \(0.001\) and a total time span of \(5\) (in arbitrary units).
We used a stochastic Heun integrator to perform the numerical simulations~\cite{kloeden2012numerical}. 
To observe the response in the non-linear dynamics domain, we introduce a perturbation of \(0.01\) on variable \(x\) at time \(0\) and then examine the effects on variable \(z\) at future time steps.
We undertake and iterate this particular system numerous times over many trajectories, assuming ergodicity of the dynamics of the system.

\begin{figure}[h]
   \centering
    \includegraphics[height=0.3\textwidth]{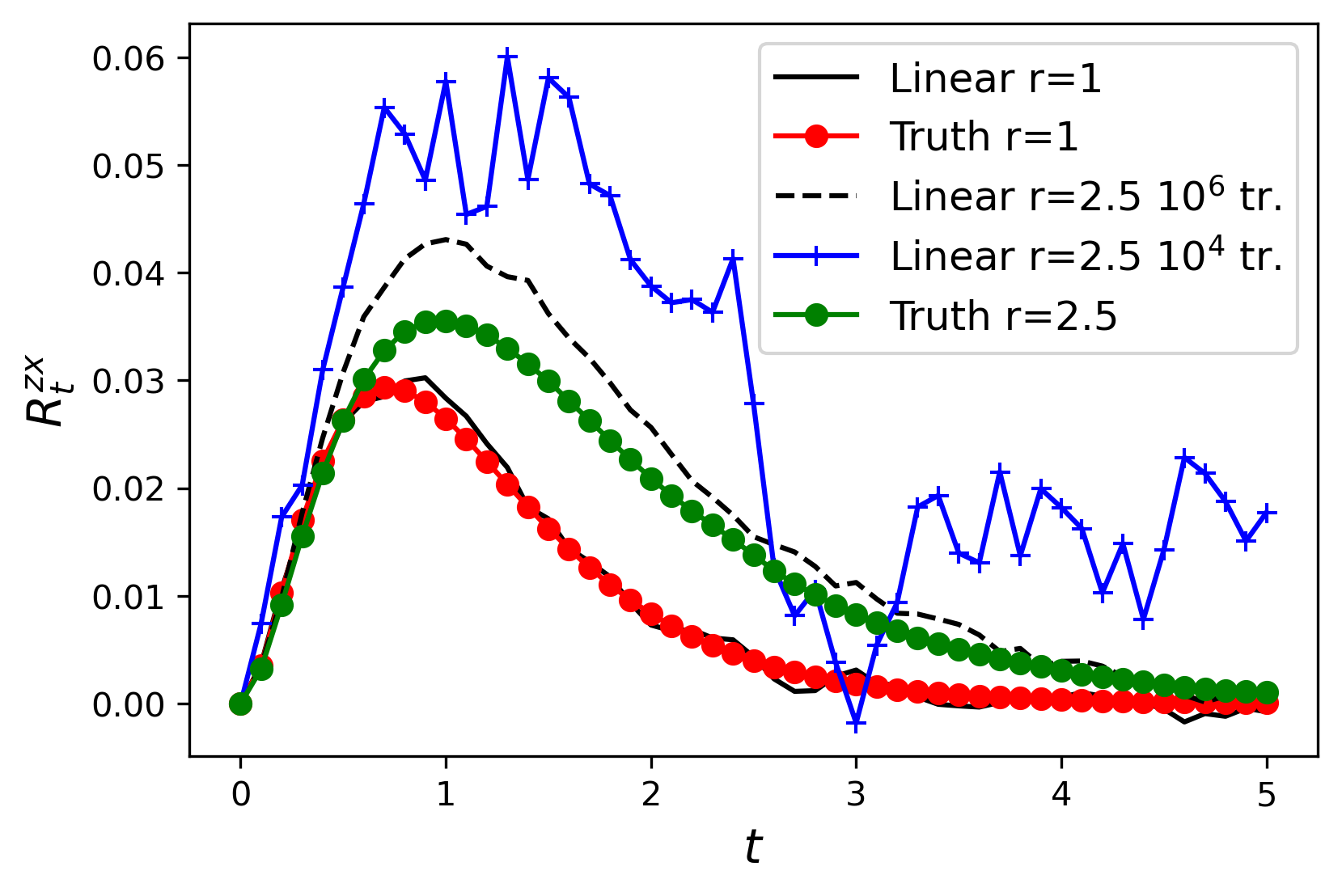}
    \includegraphics[height=0.3\textwidth]{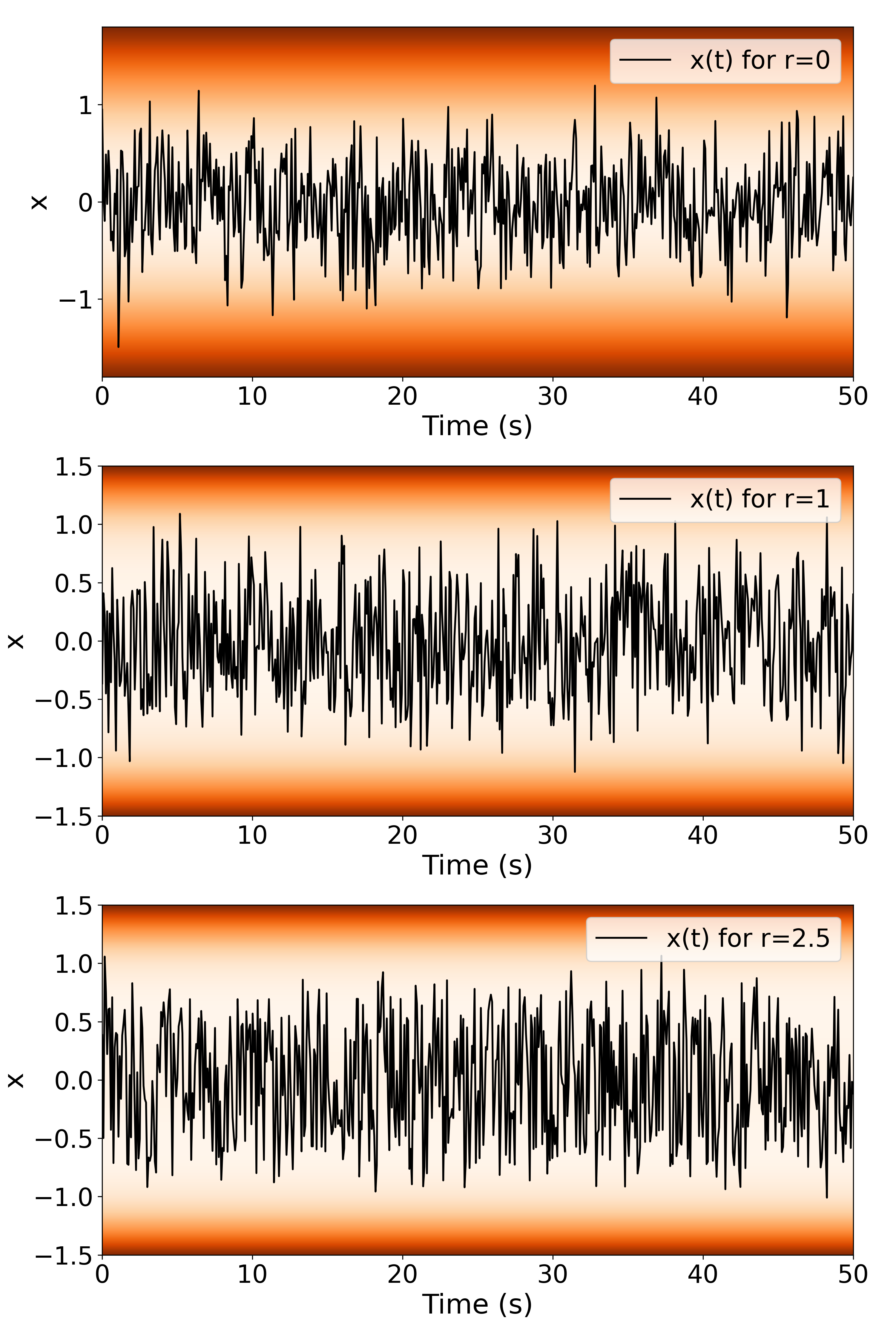}
    \includegraphics[height=0.3\textwidth]{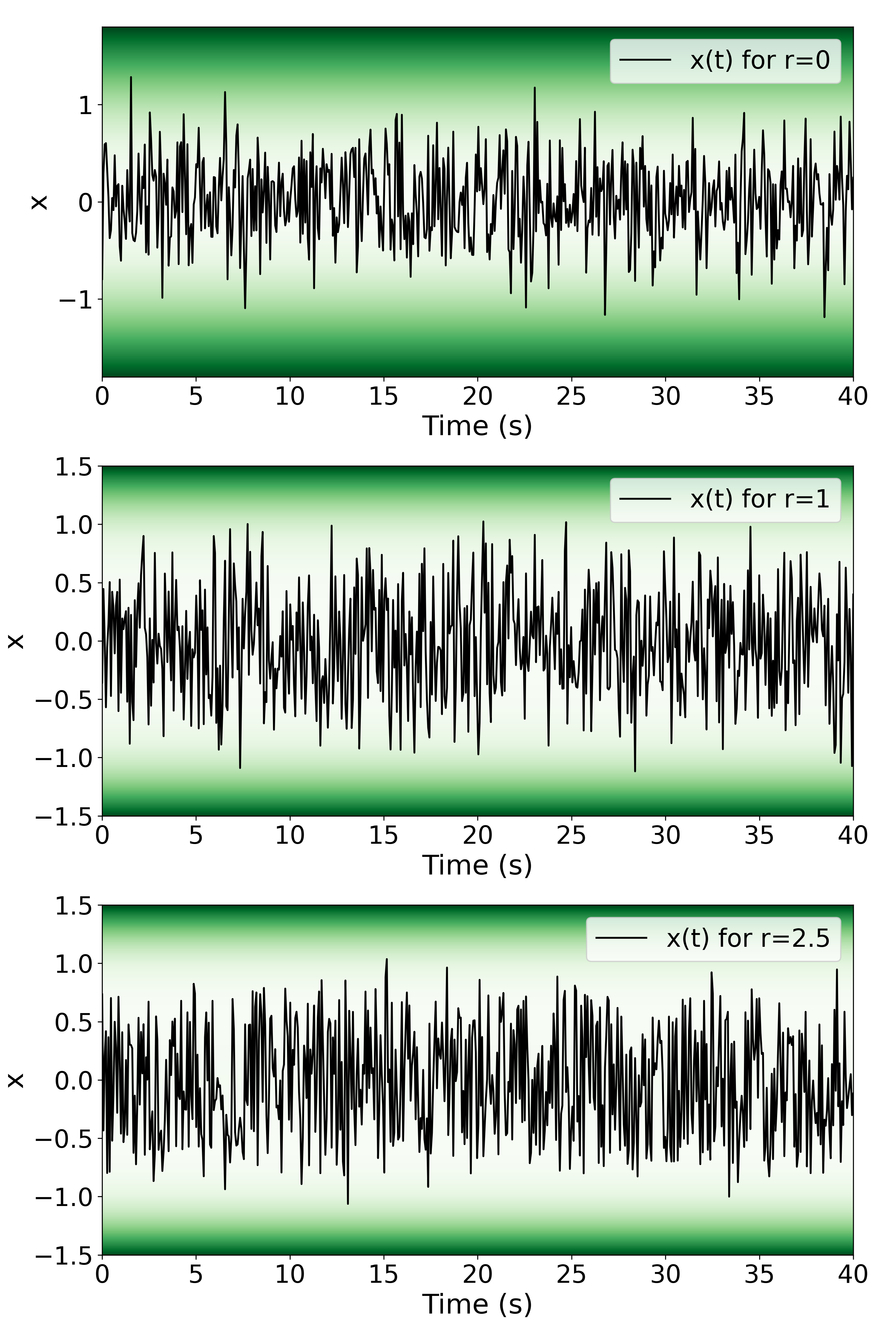}\\
    (a)\hspace{6cm}(b)\hspace{3cm}(c)\\
    \includegraphics[height=0.35\textwidth]{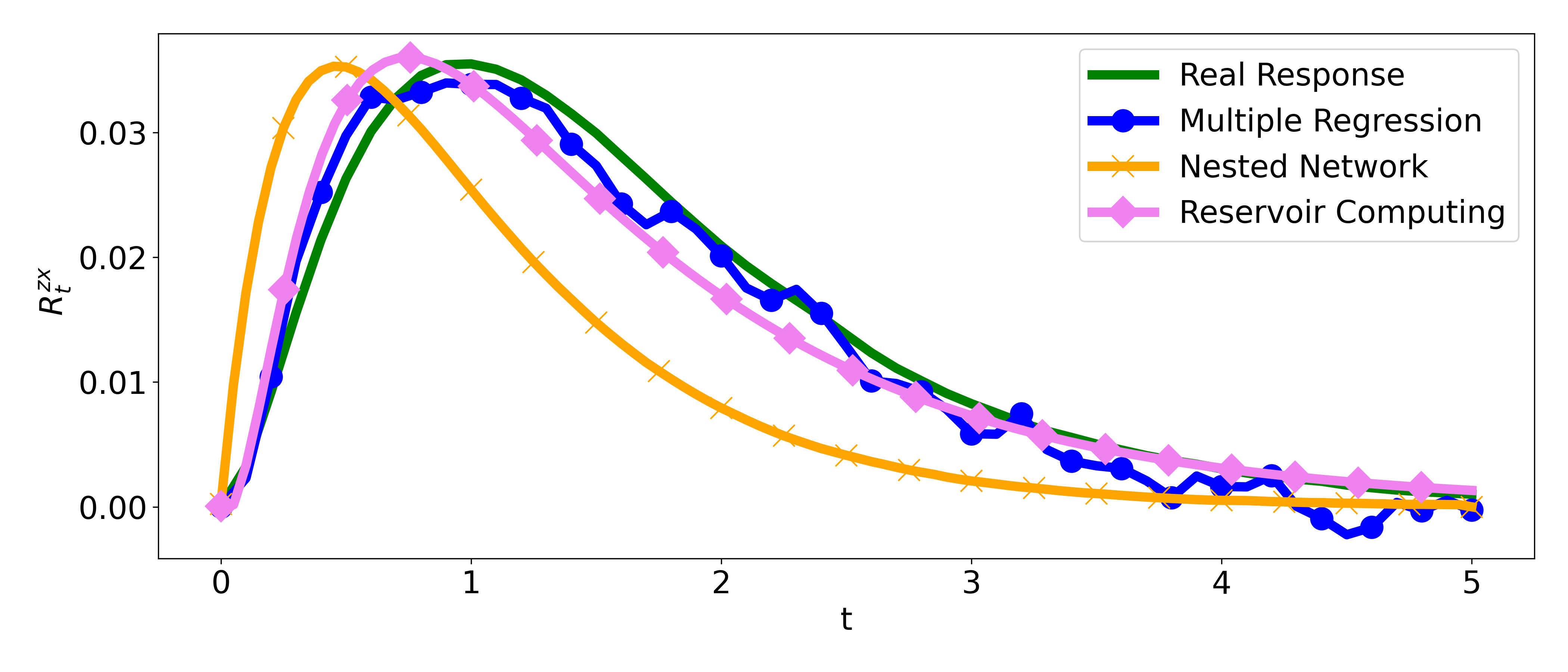}\\
    (d)

        \caption{(a) Plot of the response function \(\RR_{x\to z}\) in the non-linear dynamics for different values of \(r\);
          The response computed through definition (\ref{eq:response})  indicated as the``truth", and the linear formula based on covariance matrix (\ref{eq:resplin}) are shown.
          All the plots have been obtained by averaging over \(10^6\) trajectories, but the blue-crossed line which has been obtained with $10^4$ trajectories.
          (b)-(c): Visualisation of sample trajectories for different nonlinearity. The strength of the confining potential is qualitatively indicated by the colormap,
          with light areas corresponding to the minimum of the potential. The (b) orange figure represents the ground truth dynamics.
          The (c) green plots are obtained with the reservoir computing, see section. 
          {(d):   Comparison between the non-linear reconstruction of the response for $r=2.5$.
            Multiple Regression (blue line),  the threshold value of $\lambda$ to \(0.5 \cdot 10^{-3}\); Lasso L1 regularization was employed during the regression step, with a $\lambda$ parameter set to \(5 \cdot 10^{-6}\). The analysis was based on $10^5$ trajectories.
 Rescaled Response computed from the NN model with 34 parameters (yellow line), the amplitude has been divided by 3;
 Response obtained with Reservoir Computing (Pink line).  
  The total parameter count in the RC model was approximately 150.
   Ground truth response is given for comparison (red line).}}
    \label{fig:4}
\end{figure}
Fig. \ref{fig:4}(a) shows a comparison between the response calculated by intervention measures and the response obtained through covariance in the linear approximation. 
When \(r=1\),  an excellent agreement is found between the response computed via the general definition and its linear approximation.
In the scenario where the non-linear contribution is higher, but not too large (as in the case of \(r=2.5\)), the linearized response continues to provide meaningful insights into the
causal relationships among the system variables. It is worth emphasizing that many data must be used to avoid spurious correlations, about $10^6$ trajectories in this case.
For comparison, we show the results obtained with $10^4$ samples. In this case,
the error is much more important, and even qualitatively the response is much less clear. 
The set of equations~(\ref{eq:nl-mark}) contains a nonlinear perturbation, which is still only local in the state variables, whereas the couplings remain  linear.
This a rather general feature of physical stochastic systems, but it draws an important difference with fully non-linear chaotic dynamical systems.
The reasonable success of applying the linear approximation to this nonlinear case stems from this  structure.
We consider now different approaches to specifically address the non-linear nature of the problem, in order to be able 
to cope with more general situations of practical interest, where we are confronted just with time-series~\cite{kantz2003nonlinear}
without knowing the nature of the interaction.
These approaches are  data-driven and we want to critically assess whether they may be in some cases valuable alternatives.
In particular, we shall consider approaches with increasing the level of physical knowledge available for the modelling.

\paragraph{Non-Linear Multiple Regression}
We first make use of a regression using a nonlinear basis containing the guesses state-vector of the system;
Unlike simple linear regression, which assumes a linear relationship between variables, non-linear multiple regression allows the modelling of non-linear relationships between variables.
In doing so we are back to a linear regression setting at the expense of a potentially much larger set of candidate features, depending on the prior knowledge on the problem.
This relationship is represented by a function, which can take various forms, such as polynomial, exponential, logarithmic, or trigonometric functions, that is building upon a dictionary.
Here we choose a dictionary of size $19$ made of all $3$-variables polynomials functions up to degree $3$.
In the general case, where we are dealing with \(n\) variables up to degree \(d\), the number of coefficients is given by the formula: $ {(n+d)!}/{(n! \ d!)} $.
For the same reason as in the linear case, we use a $L_1$ regularization to enforce sparsity of the solution in order to see if we are able to recover the exact form of the potential. The choice of the
$L_1$ penalty $\lambda$ is done in a standard way by cross-validation. 
As we shall see in Section~\ref{sec:chaotic_systems},
once an optimal penalization is found, the choice of the dictionary is not key, and we would obtain similar results including a larger set of functions.
Using this approach, we are to able to reconstruct the model from data with an excellent level of accuracy, despite the stochastic nature of the system.
Let us now focus on the reconstruction of the response.
As already mentioned, within the functional formulation the dynamics is linear and we are back to the linear response setting~(\ref{eq:resplin}) without approximations but 
in a much larger space.  

As for the linear case, we may regress the Markov operator and use iterations of this estimate operator to compute the generalized response. Hence,
regression is performed between $t$ and $t+1$, and the linear response for arbitrary time horizon \(\tau\)
is obtained by taking powers of the inferred Markov operator, where we can selectively focus only on the first three rows,
which refer to the variables \(x\), \(y\) and \(z\) and ultimately define the state of the system. 
Capturing the response requires summing over all the variables that depend on \(x\),
thus contributing to the dynamics of the system. In particular,
the polynomial features \(x\) and \(x^3\) play a central role and we cannot neglect the corresponding response generated by their influence.
As can be seen from \ref{fig:4}.c, this method lead to an accurate representation of the response curve, hence showing how the linear response theory
can be adapted to nonlinear systems.

\paragraph{Nested Neural Network}
\begin{figure}[h]
    \noindent
    \centering
    \includegraphics[width=0.4\textwidth]{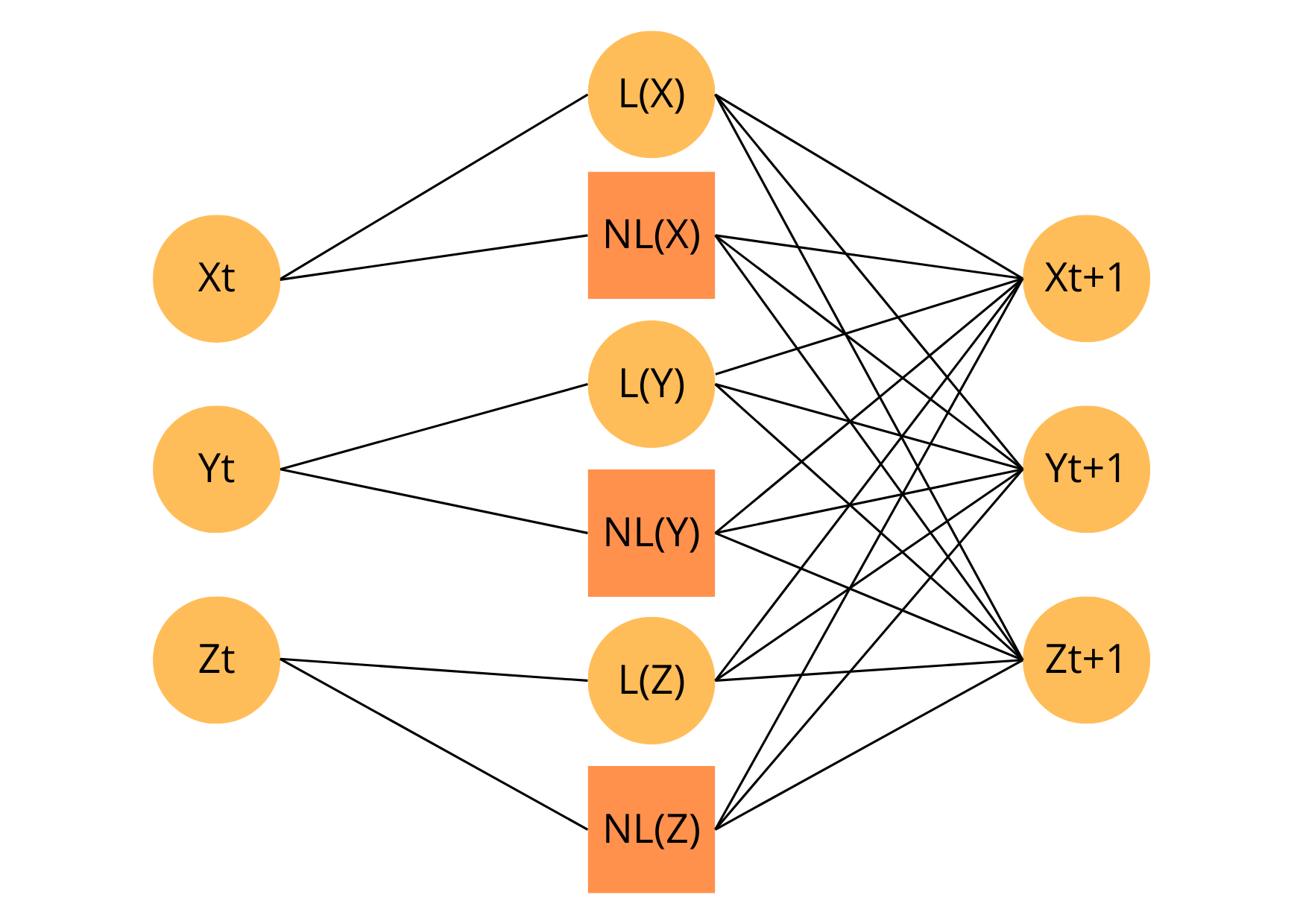}
    \includegraphics[width=0.5\textwidth]{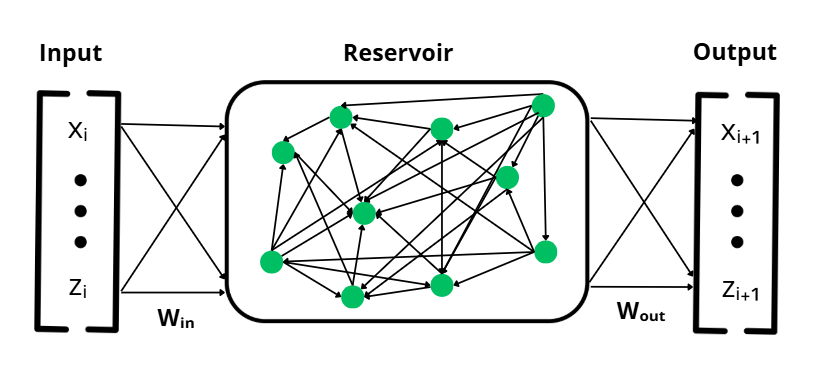}
    \caption{Left: Architecture of the Nested Neural Network in which each node representing the variable is connected to a linear block and a non-linear block. The linear block is a single node, while the non-linear block is characterized by two labels with two neurons each. Everything is fully connected to the output label. The network is characterized by 34 parameters. Right: architecture of a reservoir computing. Only output weights are trained. Internal weights $W$ are kept fixed.}
    \label{fig:8}
\end{figure}
Before going for a pure data-driven approach, we have also developed an original approach particularly designed to sort out causal relations among variables, given some prior form of
the model. The approach is therefore data-driven but takes into account some physical considerations, leading to an efficient model in terms of need of data.
That may be relevant when a pure data-driven approach becomes too greedy in data to correctly build a model of nonlinear systems~\cite{bucci2023curriculum}.
The architecture proposed is sketched in Fig. \ref{fig:8}.
The model is formulated first by treating the linear and nonlinear part of the dynamics separately in terms of two distinct channels.
Again the focus is on predicting the future state of a system defined by three dynamical variables — $x, y$, and $z$ — observed at a given time step t,
and estimating their respective values at the following time step t+1.
While in principle we do not make any assumptions on the amplitude of the nonlinearity, yet  we consider implicitly that the nonlinear couplings are weak.
 The model processes the input through a dual-pathway design, combining both linear transformations and  nonlinear mappings via nested sub-networks. 
 This hybrid approach leverages the interpretability of linear models along with the expressive power of neural networks to capture complex temporal dynamics.
At each time step, the system receives a three-dimensional input vector composed of the current values of the variables $x, y$, and $z$. Each of these variables is then processed through two distinct computational channels:
(i) Linear Channel: the first path applies a simple linear transformation to the input variable. This channel serves to retain the original signal in a scaled form, ensuring that straightforward linear trends or proportional changes are preserved throughout the network. Each variable is multiplied by a learnable coefficient, providing a direct and interpretable signal flow to the final output layer.
(ii) Nonlinear Channel (Nested Neural Network): In parallel, each variable is passed through an independent, nested neural network. These sub-networks comprise multiple hidden layers with nonlinear activation functions, enabling the modelling of complex and potentially nonlinear temporal relationships. Each nested network is dedicated to a single variable.
The outputs from both the linear and nonlinear channels for each variable are then combined — typically through concatenation or summation — and forwarded to the final layer of the network.
This final layer is composed of three neurones, each responsible for predicting the value of one of the original variables at the next time step.
This dual-channel strategy offers several advantages. The linear pathway preserves a direct, interpretable signal, which can be important in domains where understanding the contribution of
individual variables is necessary.
Meanwhile, the nested neural networks allow the system to go beyond simple trend following and instead capture latent interactions,
nonlinear dependencies, and dynamic behaviours that would be inaccessible to purely linear models.
This architecture is modular. Each variable is handled through its own nested sub -network, allowing for isolated tuning, parallel processing, and potential reuse across tasks or models. 
The interest of this approach is its compatibility with explainability tools. The clear separation between linear and nonlinear components allows analyse the distinct
contributions of each path to the final output, gaining insight into which aspects of the input are being handled linearly and which are subjected to more complex transformations.

The reference response to be compared with is computed using the standard recipe:
the two systems are initialized with the same initial conditions and the value of \(x\) is shifted by \(1\%\) in the second system. In the next steps, we calculate the average variation of
the variable \(z\) by evolving the system through the Nested Neural Network model. The causality can be determined quantitatively from the level of the response which is observed.
Interestingly, even a small yet physically oriented model is able to provide decent results,
as highlighted in Figure~\ref{fig:4}(d), where we show the results of the response obtained with a small network including only 34 parameters. 
It is remarkable that we have obtained correct qualitative results even with small training datasets.
\paragraph{Recurrent Network Model: Reservoir Computing}
Here we turn to a purely data-driven approach without any implicit assumption on the process. 
The idea is to learn a  model of the process on which the interventional causal experiment can be explicitly performed
in the form of~(\ref{eq:resp}). 
Many architectures/models can be considered for this kind of tasks. 
When one considers a dynamical system, a sensible approach is to use a recurrent neural network (RNN)~\cite{vlachas2020backpropagation}. 
A specific network which has been found particularly
successful for nonlinear dynamical systems is the Reservoir Computing (RC). Over the years, RC has attracted considerable attention due to its ability to solve problems in time series prediction, classification and control, providing reliable results with minimal parameter tuning. Despite its conceptual sophistication, the practical utility of RC has solidified its role as a central tool in computer modeling~\cite{schrauwen2007overview,pathak2018model,storm2022constraints,storm2024finite}.
 RC was introduced by models such as Echo State Networks (ESNs) and Liquid State Machines (LSMs) and simplifies the training process by using a randomly initialized, fixed recurrent layer, the so-called reservoir. Instead of training the entire network, as is the case with traditional RNNs, only the output layer is optimized, making RC extremely efficient and scalable. 
RC has been shown to be particularly effective when dealing with systems characterized by a completely Markovian nature, where transitions depend only on the current state and require no memory of previous time steps. In such contexts, traditional RNNs often lead to inefficient learning as they attempt to model unnecessary temporal dependencies that are irrelevant to the problem at hand. The core concept of RC is to project a non-linear system into a higher dimensional space in which its dynamics become approximately linear. This transformation allows the application of linear regression techniques, often augmented with Lasso L1 regularization, to manage sparsity and improve model generalization. Unlike standard RNNs, which require iterative and complex weighting updates, the RC reservoir remains fixed, significantly reducing computational complexity while preserving the expressive power of the system. This approach generalizes the fixed dictionary based regression by allowing the features to be learned.
 RC has proven to perform exceptionally well in a variety of applications, even when recurrent dynamics are disabled, as memory is not always required for certain problems. This black-box approach combines efficiency with versatility and is therefore particularly suitable for systems with minimal or uniform causal dependencies over time. 
 
In the problem under consideration, a three-dimensional input space was mapped to a 50-dimensional feature space. 
Reducing the dimensionality to less than 50 dimensions led to a small shift of the response curve in Fig.\ref{fig:4}(d) to the left, but did not significantly change the overall results, and the agreement between the ground truth and the response predicted by the network is quite good. From a more qualitative point of view, we show in Fig.~\ref{fig:4}(b-c) the signals produced by the original stochastic equations (in orange) and by the RC, for different non-linearity. It can be appreciated that trajectories appear very similar from a statistical point if view.
This approach achieved highly accurate responses that were consistent with prior methodologies. 
These findings highlight the versatility and efficiency of RC in addressing a wide range of computational challenges.

Summarizing the results of this section, when the system is weakly non-linear, insights on  causal relationships may be found already with the linear approximation.
Then for more strongly non-linear systems, a kind of physics-informed machine learning approach may be used based on a regression of nonlinear functions,
allowing us to compute exactly the response in the linear framework.  
The main limitation comes from the size of the training set. When a large data set is given, the most efficient model is the RNN approach.
Then in the situation of a limited dataset, a trade-off can be found with a predefined feature model
by adjusting the size of the dictionary. Yet when data are too scarce, the nested model which has fewer parameters can still give a good qualitative picture.

\section{Causality in fully non-linear systems}\label{sec:chaotic_systems}

We now consider the Lorenz '63 system, which exhibits a highly nonlinear and chaotic behaviour. This system's evolution equation reads
\begin{equation}
\left\{\ \begin{array}{l}
\dot{x}=\sigma(y-x) \\
\dot{y}=r x-y-x z \\
\dot{z}=x y-b z 
\end{array}\right.
\label{eq:eq-Lorenz}
\end{equation}
where $\sigma$ represents the Prandtl number, $r$  the Rayleigh number and $b>0$ is a parameter linked to the ratio of the convective rolls.
With $\sigma=10$, $b=\frac{8}{3}$, $r=28$ the system display fully chaotic behavior~\cite{lorenz1963deterministic,ott2002chaos}. 
This special case fall into the general class of dynamical system characterized by $d$ scalar variables $\mathbf{x} = \left(x_0, x_1, \dots, x_{d-1} \right)$, given by the relation
\begin{equation}
    \dot{\mathbf{x}} = \mathbf{f}(\mathbf{x}),
\end{equation}
for a vector valued function $\mathbf{f} \colon \mathbb{R}^d \to \mathbb{R}^d$.

As for the stochastic systems, we consider the system to be fluctuating around an invariant measure and are interested in the study of
a data-driven interventional approach for computing the response functions
\begin{equation}
    \RR_{j\to i}(\tau) = \frac{\overline{\delta x_i(t + \tau)}}{\delta x_j(t)},
\end{equation}
where $\overline{\cdot}$ represents the ensemble average and $\delta x_i$ denotes the perturbation that $x_i$ undergoes when $x_j$ is perturbed with $\delta x_j$.

In order to do this, we assume  having access to a long trajectory $(\mathbf{x}(t))_{t \in [0, T]}$, while $\mathbf{f}$ is kept hidden to us.
Since we are not able to influence any of the variables of the system, we resort to a data-driven method for approximating the perturbed trajectories.
First, we try to predict $\mathbf{f}$ based on the given $(\mathbf{x}(t))_{t \in [0, T]}$. To this purpose, we train an adequate Neural-Network with the accessible data.
In order to approximate its response functions we have implemented two separate methods of approximating $\mathbf{f}$:
\begin{enumerate}
    \item[(a)] Multi Layer Perceptron (MLP)~\cite{rosenblatt1958perceptron} made up of a $3$ layers feedforward neural network with 50 neurons in each hidden layer and LeakyReLU(0.1) as an activation function. This model has been trained with up to $10^6$ samples.
    \item[(b)] Sparse Identification of Nonlinear Dynamical systems (SINDy) \cite{Brunton_2016} implemented using the PySINDy python package \cite{desilva2020,Kaptanoglu2022},
      which is based on performing a nonlinear regression on a basis of functions. 
    This model has been trained on $10^4$ samples;
\end{enumerate}
Then, as for the reservoir computing, we can perturb many trajectories of our synthetic model  to numerically compute the response based on the simulated modified trajectories.
This allows us to make interventions on the system and produce an arbitrary number of samples to compute accurately the statistics based on the learned model.

We will compare the response results obtained via these approaches also with those of the linear approximation which consists in computing the response based on the covariance matrices.

Following the results obtained in the previous section, a simple data-driven approach like the MLP appears adequate.
We expect the nonlinear chaotic systems to perform like a Markovian one~\cite{vulpiani2010chaos}, and therefore recurrent networks are not needed. Nonetheless,
we have thoroughly checked that the results remain qualitative the same using a RNN, so that present
results are to be considered general.
For a more physics-oriented approach we have used SINDy, which is conceptually very close to the nonlinear regression previously used for stochastic processes. In SINDy,
complex nonlinear dynamics are represented as a linear combination of simpler nonlinear functions. These functions (feature space) typically include terms inspired by the
underlying governing equations, allowing the algorithm to incorporate relevant physical information \cite{loiseau2018constrained}. 
In a nutshell, we may approximate a function as 
$ \mathbf{f(x)} \approx \sum_{k=1}^p \alpha_k \Xi_k({\bf x}) $, where the dictionary is $\Xi({\bf x}) = [{\bf 1\; x\; x^2 \; \ldots x^k \; \ldots sin(x)\; \exp(x)}\; \ldots ]$,
and $\alpha$ is a sparse vector with only some nonzero terms.
In our case, we have included in the dictionary polynomials {up to second order},
trigonometric and exponential functions. Combining the training with the sparse penalisation\cite{Brunton_2016} , we obtain the coefficients in front of each feature.
To this end we minimize the Loss function ${\bf \alpha} = \argminB_{{\bf \alpha}^\prime} || \dot{{\bf x}} - \Xi({\bf x}) {\bf \alpha}^\prime ||_2 +\lambda || {\mathbf \alpha}^\prime ||_1$.


\begin{figure}[h]
    \centering
            \includegraphics[width=0.32\textwidth]{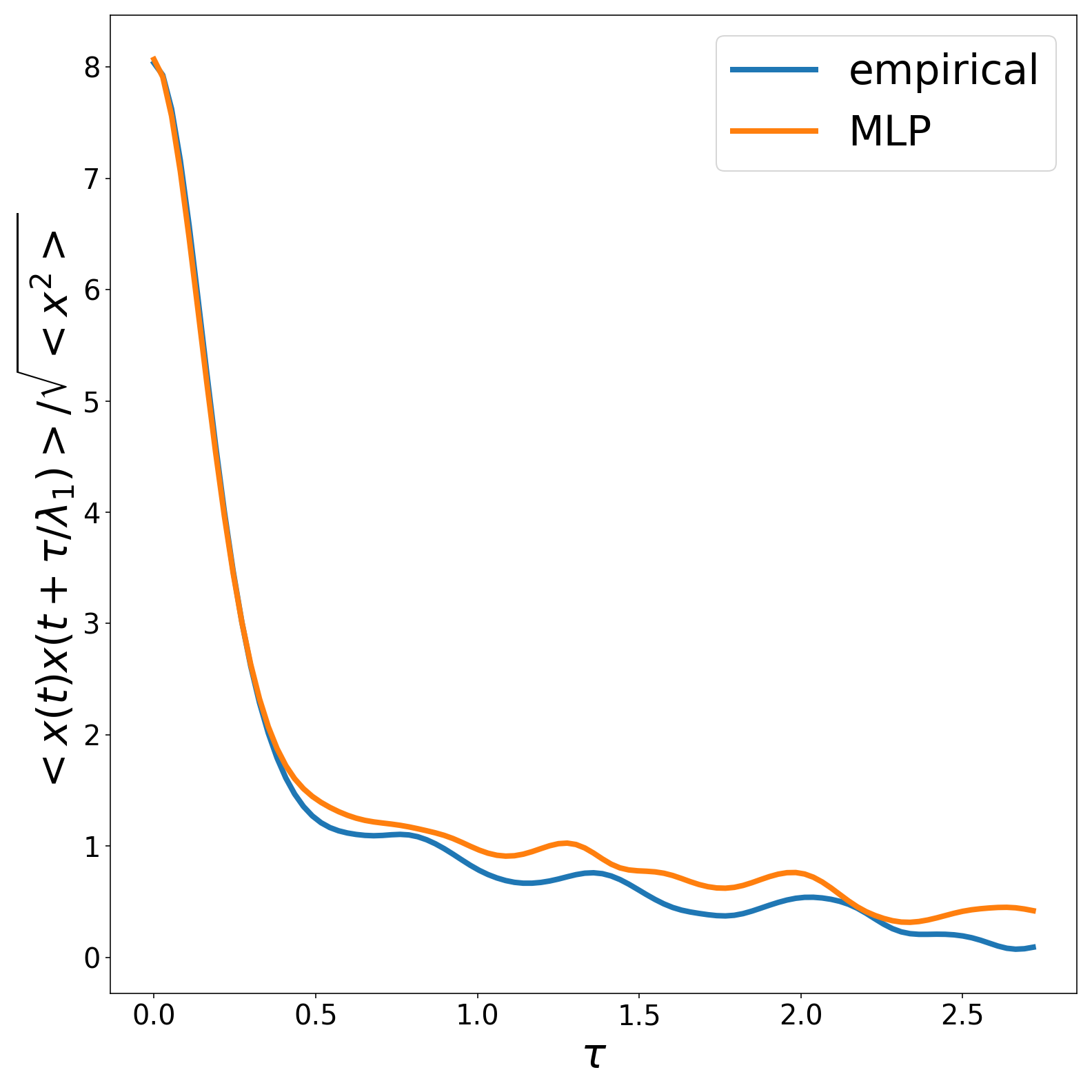} 
            \includegraphics[width=0.32\textwidth]{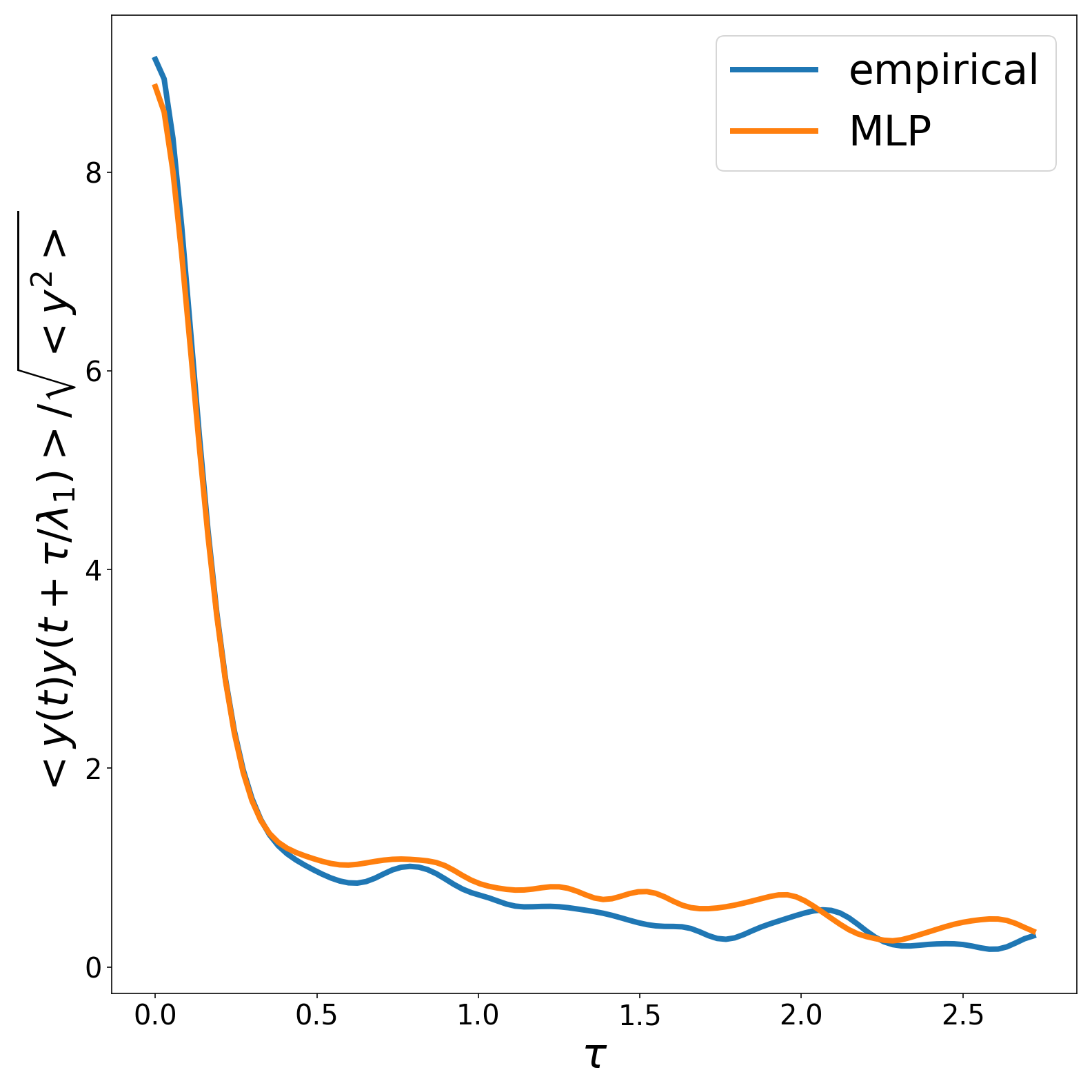} 
            \includegraphics[width=0.32\textwidth]{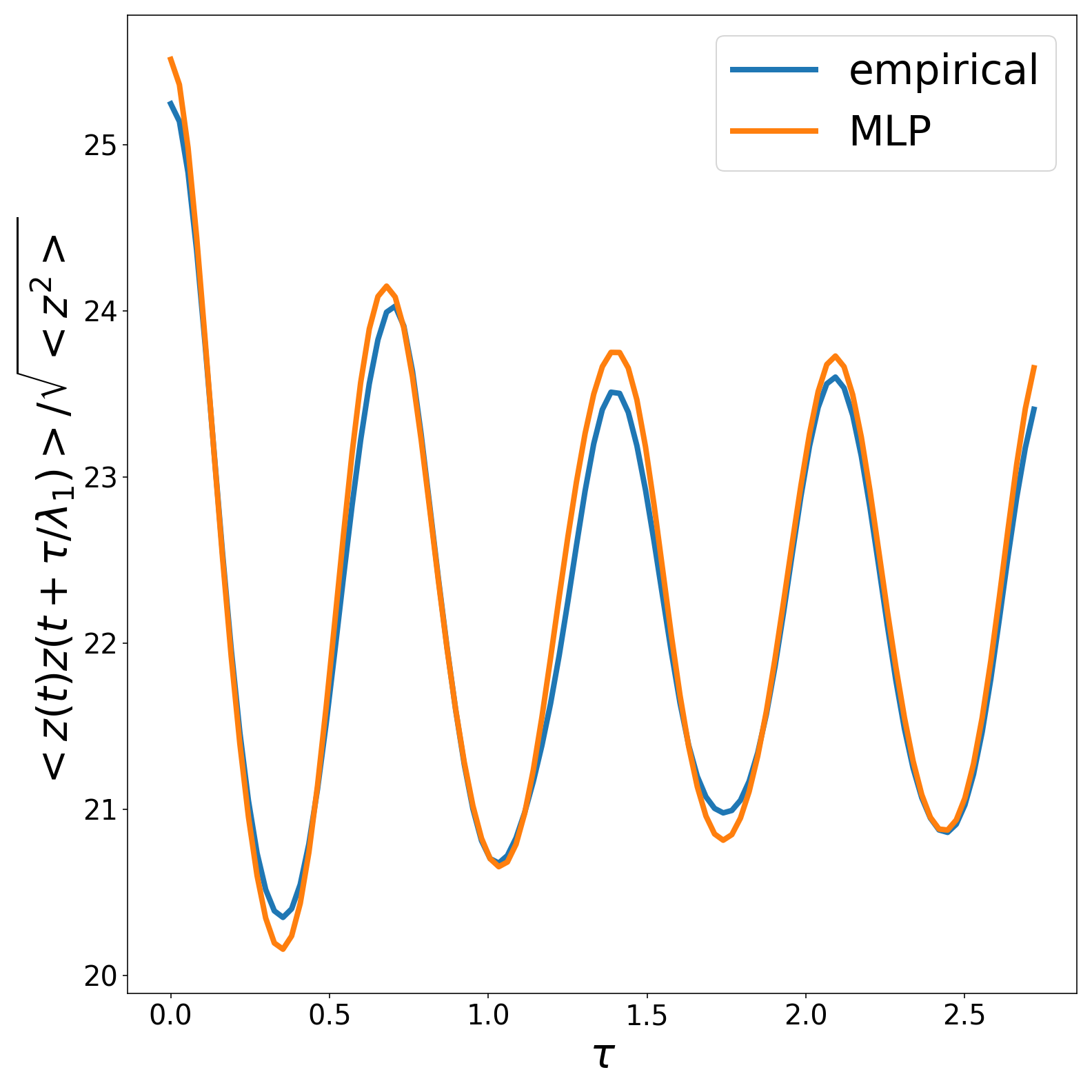} 
            \caption{Normalized auto-correlation function $<x_i(t) x_i(t+\tau)>$ for the MLP model compared against that computed through direct integration of the Lorenz system Eq. (\ref{eq:eq-Lorenz}) (blue line), and the SINDy model (green line). Lorenz and SINDy are indistinguishable. The time is expressed in terms of the Lyapunov time based on the maximum Lyapunov exponent.}
    \label{fig:auto}
\end{figure}
\begin{figure}[h]
    \centering
            \includegraphics[width=0.4\textwidth]{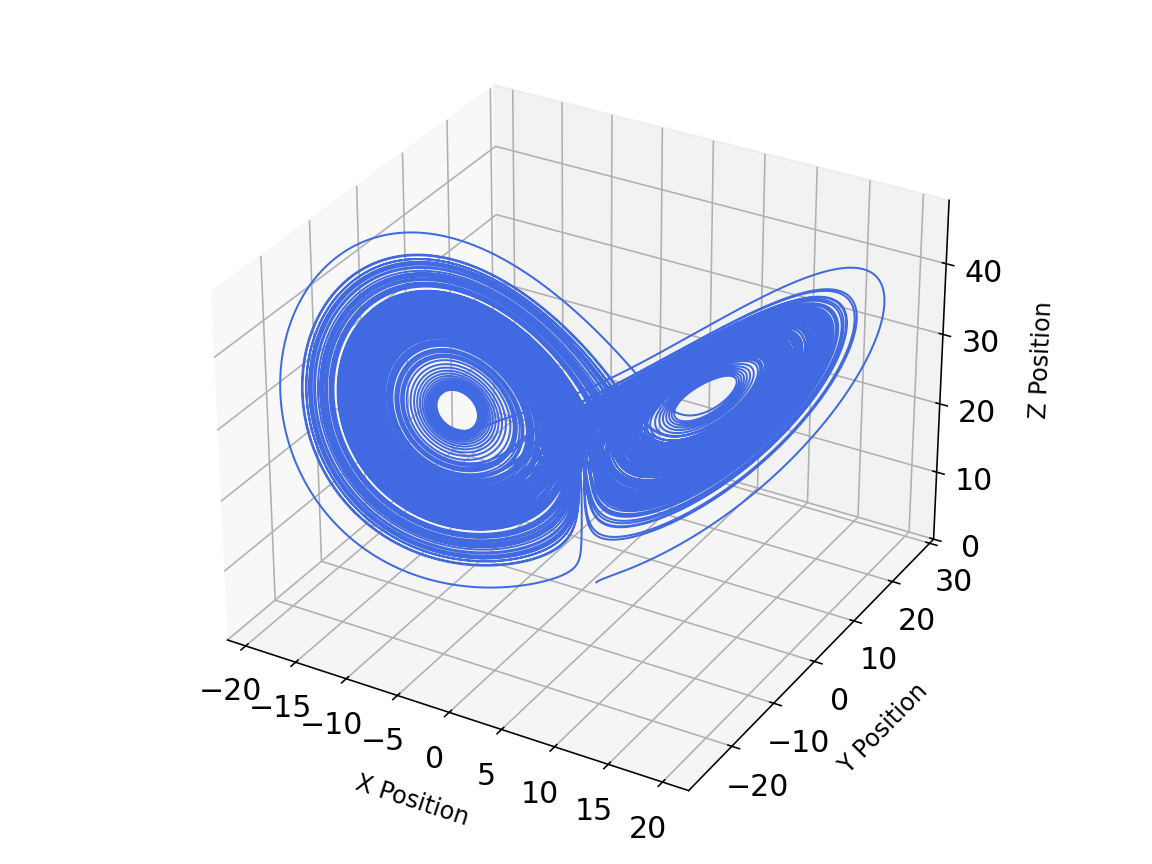}
            \includegraphics[width=0.4\textwidth]{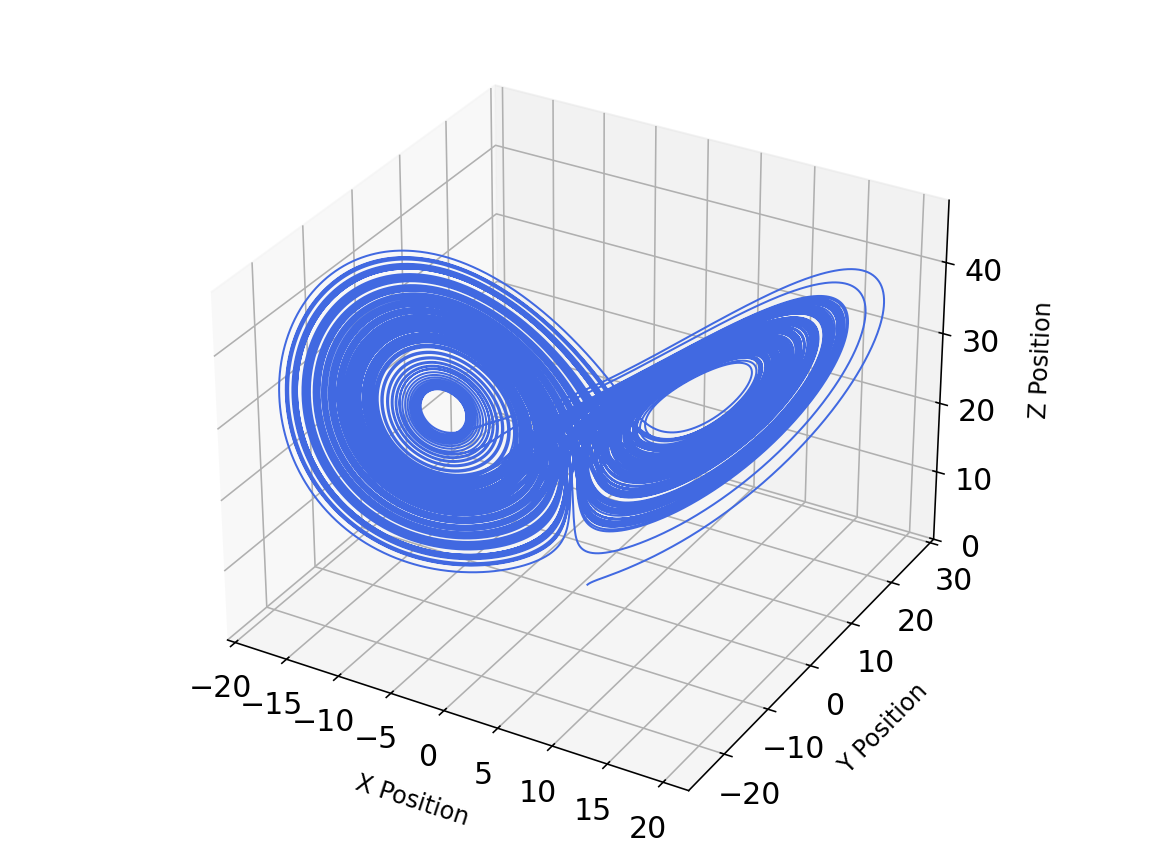}            
            \caption{ Trajectory obtained through the numerical integration of the Lorenz system (Left) compared with that generated using an MLP NN (right). }
    \label{fig:attr}
\end{figure}

First, we look at the quality of the reconstruction of the phase-space dynamics by the models. Through the sparse identification we are able to find the coefficient of the
Lorenz model basically exactly.
In Fig. \ref{fig:auto}, we show the auto-correlations for each variable and we can see that it is not possible to distinguish the True model from the SINDy one.
Looking at the results obtained with the MLP, it is found that the agreement is excellent. If some small difference may be observed, it is worth emphasising that
all the observables are equivalent from a statistical point of view. 
Additional discussion concerning the correlation can be found in appendix.
Pursuing the comparison of the dynamics, we show in Fig. \ref{fig:attr} the typical attractor obtained using the true model or the MLP; It is evident that the classical butterfly Lorenz
attractor is accurately reconstructed.
It is worth noting that these particularly clean results  have been obtained for the MLP thanks to a large data-set training; similar results have been obtained also with smaller chunks of data, with a little more of statistical noise. For the present proof-of-concept goal, and without loss of generality, we stick to the large dataset, to avoid possible issues related to the lack of training. 

\begin{figure}[h]
        \includegraphics[width=0.35\textwidth]{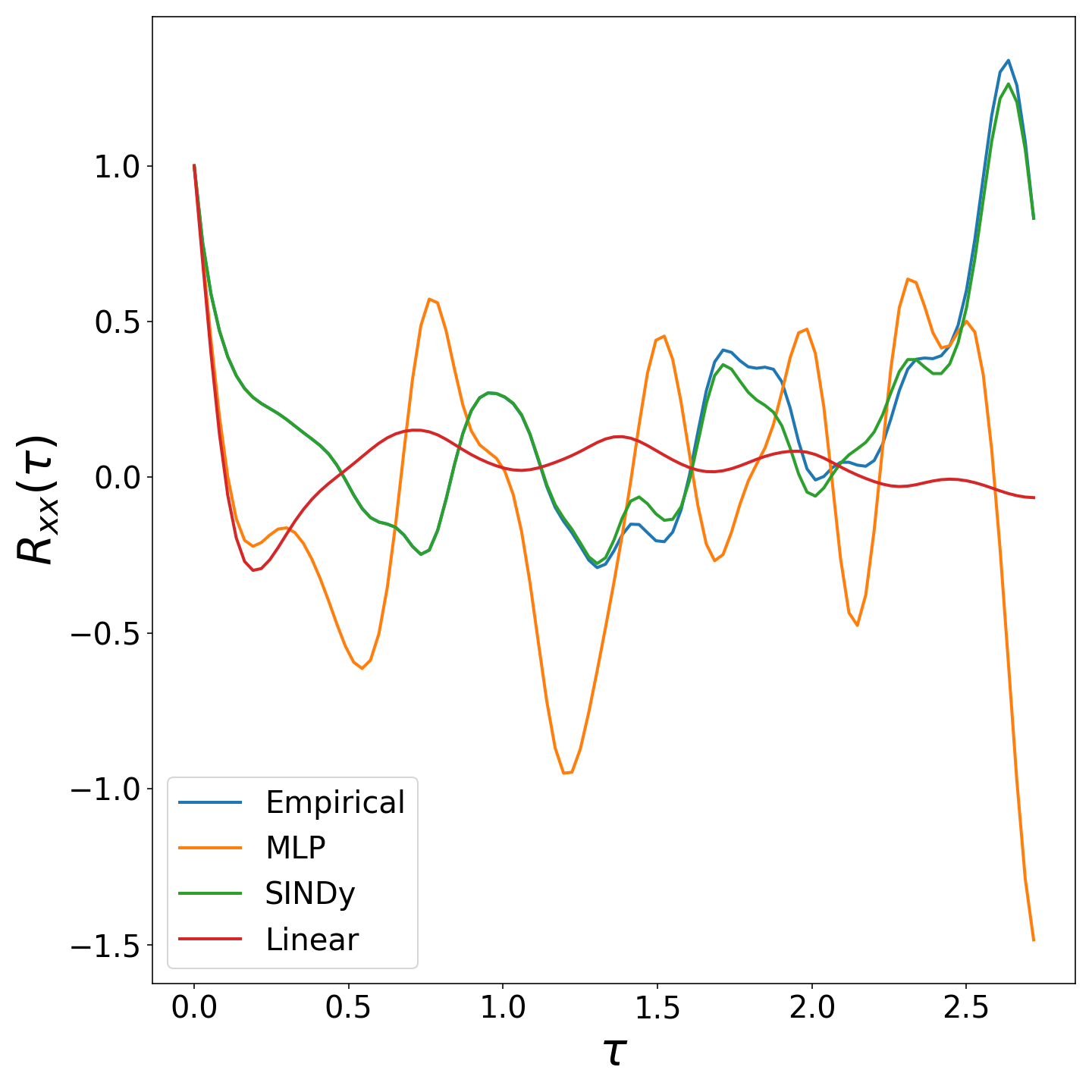}
               \includegraphics[width=0.35\textwidth]{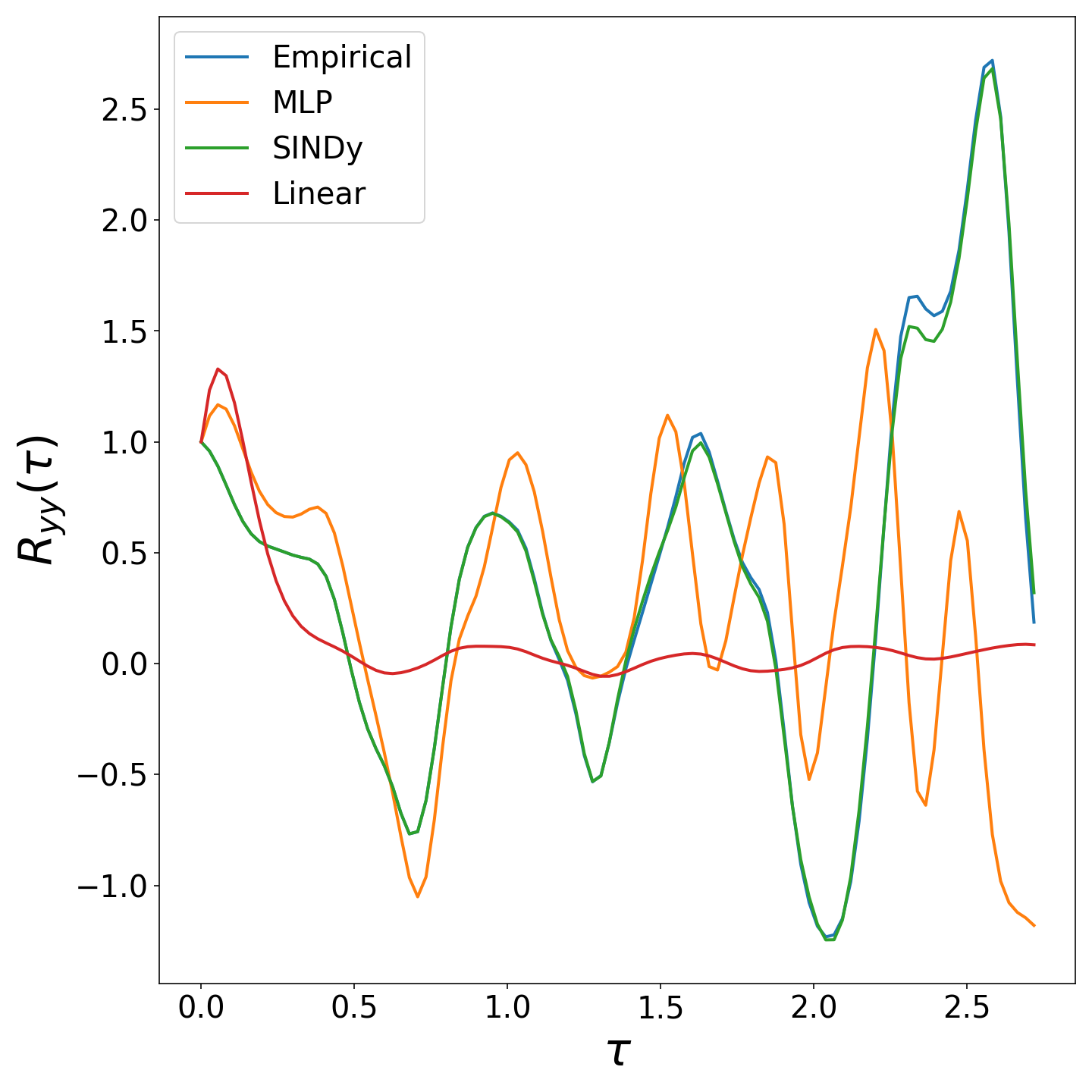}\\
       \includegraphics[width=0.35\textwidth]{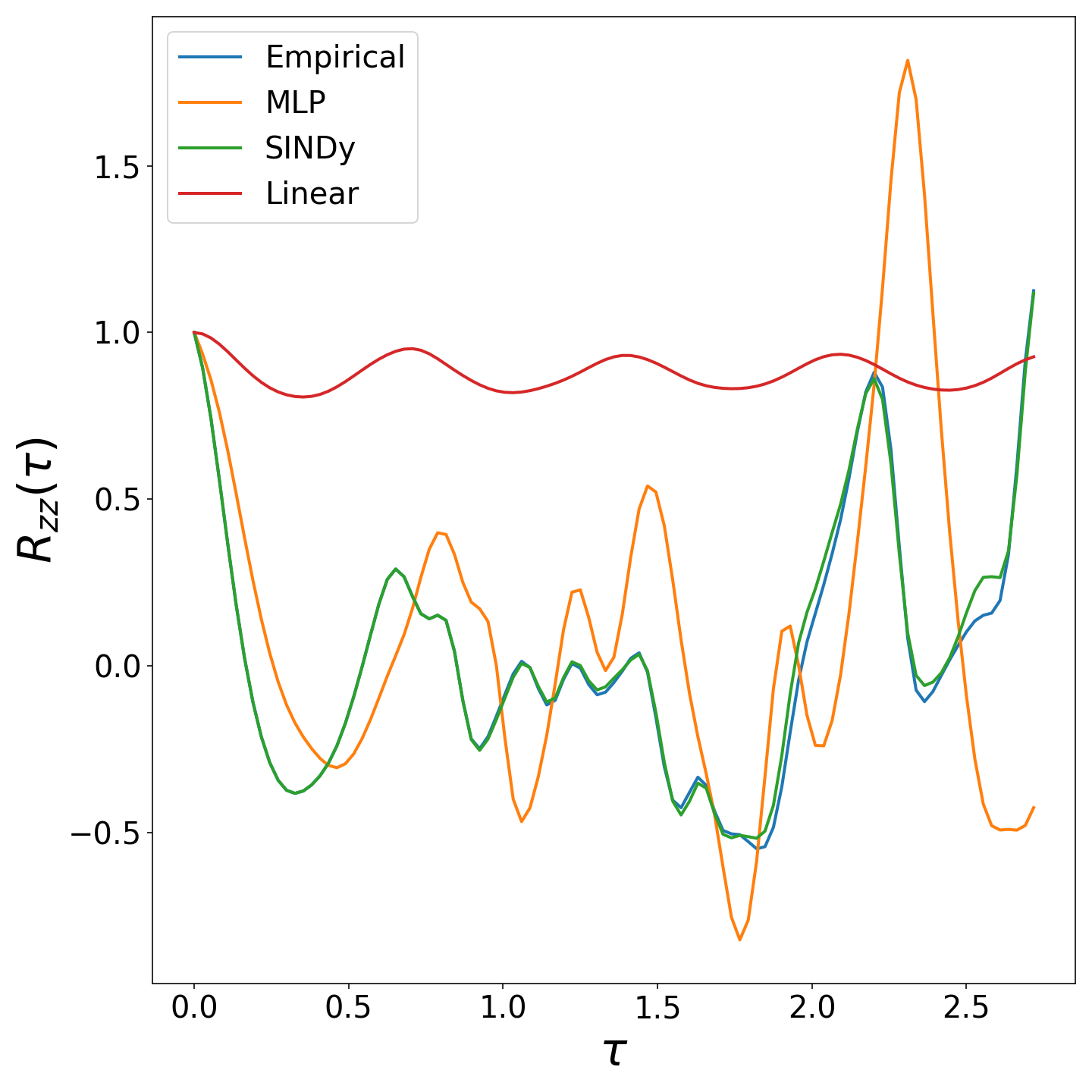}
       \includegraphics[width=0.35\textwidth]{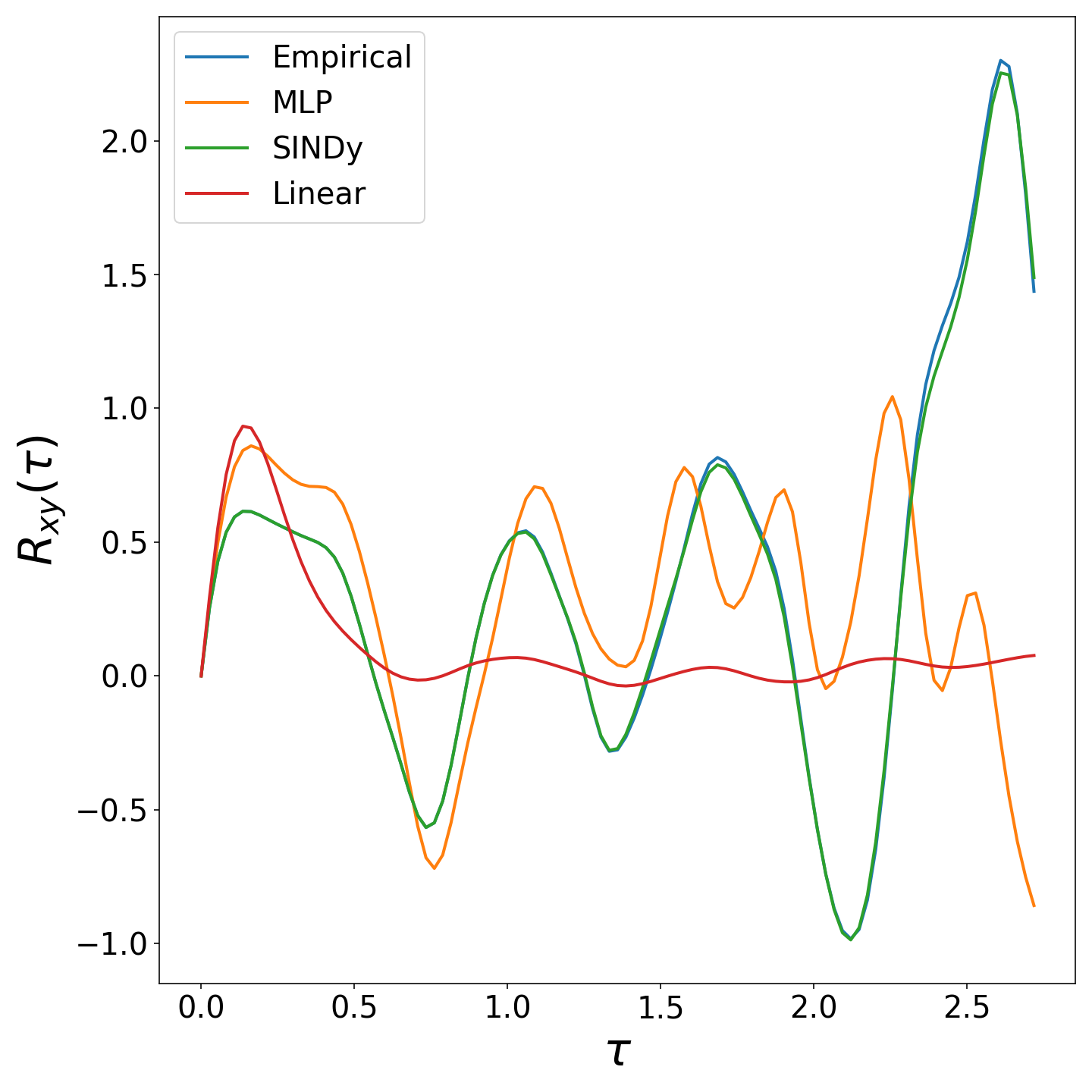}
    \caption{The response functions $\RR_{i\to j}(\tau)$ for $i,j = {x,y,z}$, obtained numerically with a sample size of $5000$, using the two methods presented above, the empirically obtained response functions using a Runge-Kutta method of order 4 for integrating the dynamical system \eqref{eq:eq-Lorenz} and the response functions predicted by the linear response theory. The unperturbed trajectory is computed using RK4. The time-scale is normalized by the Lyapunov time $1/ \lambda \approx 500$ units, $\lambda$ is the maximum Lyapunov exponent. }
    \label{fig:lor-resp1}
\end{figure}

We consider now the analysis of the causality through the response. In the Lorenz model, there are different couplings between variables, such as linear and nonlinear. Moreover, there are self-linear interaction for all variables. It is interesting to analyse the results disentangling the various couplings.
Figure~\ref{fig:lor-resp1} presents a comparison of the numerically obtained response functions for the linear interactions. 
First, It is not surprising that the response predicted for the linearized system performs poorly, due to the strongly nonlinear character of Lorenz '63. It is still interesting to remark that the linear approximation provides poor qualitative predictions even when  considering just the linear part of interactions. When the model is fully nonlinear, the dynamics on the attractor cannot be trivially reduced to its linear part.
As a consequence of the accurate reconstruction of the original model, SINDy performs outstandingly, the response functions obtained in this way  overlap with the empirically obtained response for most of the time. Moreover, the MLP approach outperforms the pure linear model, and in particular reproduces very well $\RR_{y\to x}$.
Both $\RR_{x\to x}$ and $\RR_{z\to z}$ responses are also correctly reproduced by the MLP except over long time horizon.
This limitation arises due to chaos, which inherently prevents long-term predictions when even small, finite discrepancies exist in the model.
However, we can observe that both global qualitative and quantitative prediction are good, the main frequencies of the response being detected.
\begin{figure}[h]
                \includegraphics[width=0.35\textwidth]{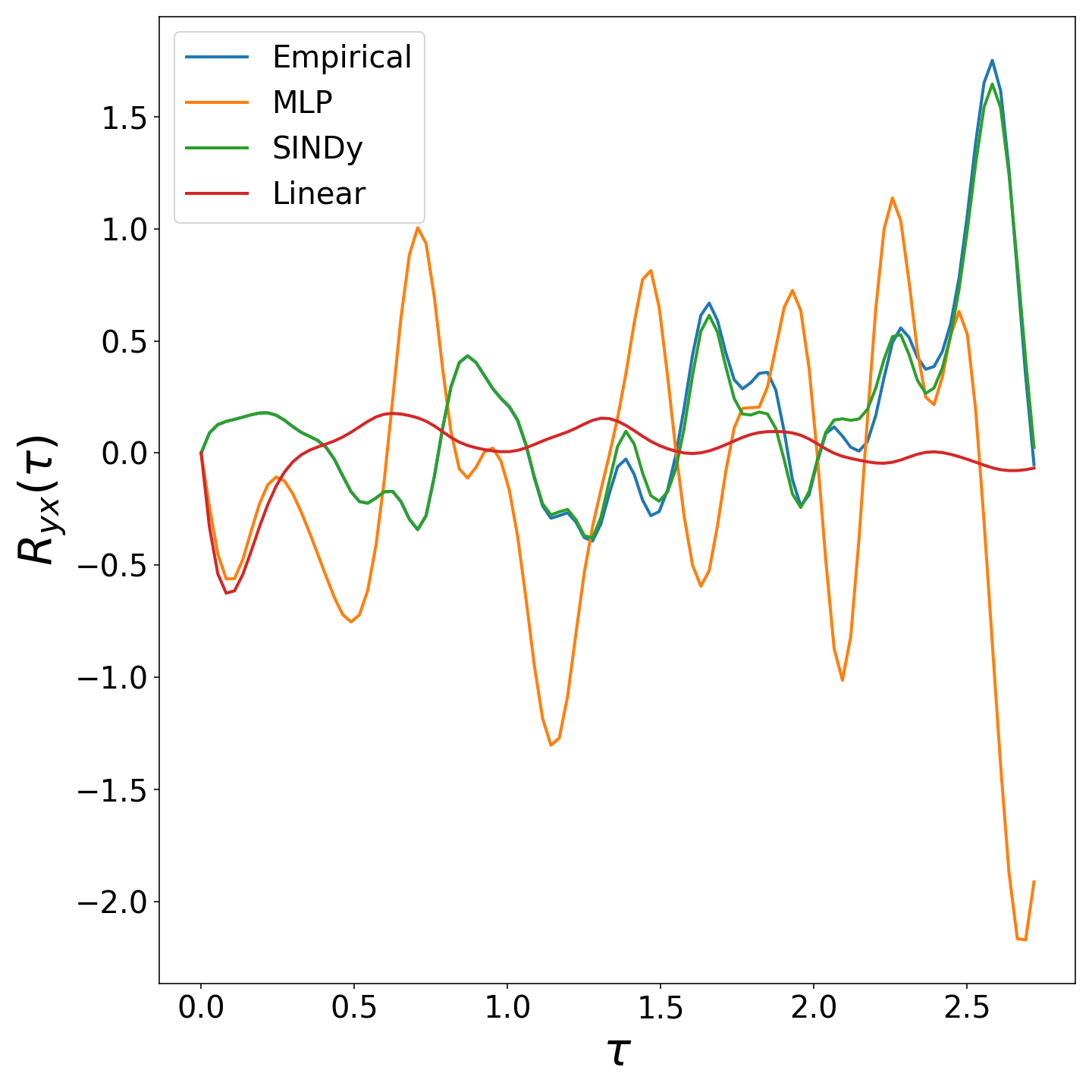}
       \includegraphics[width=0.35\textwidth]{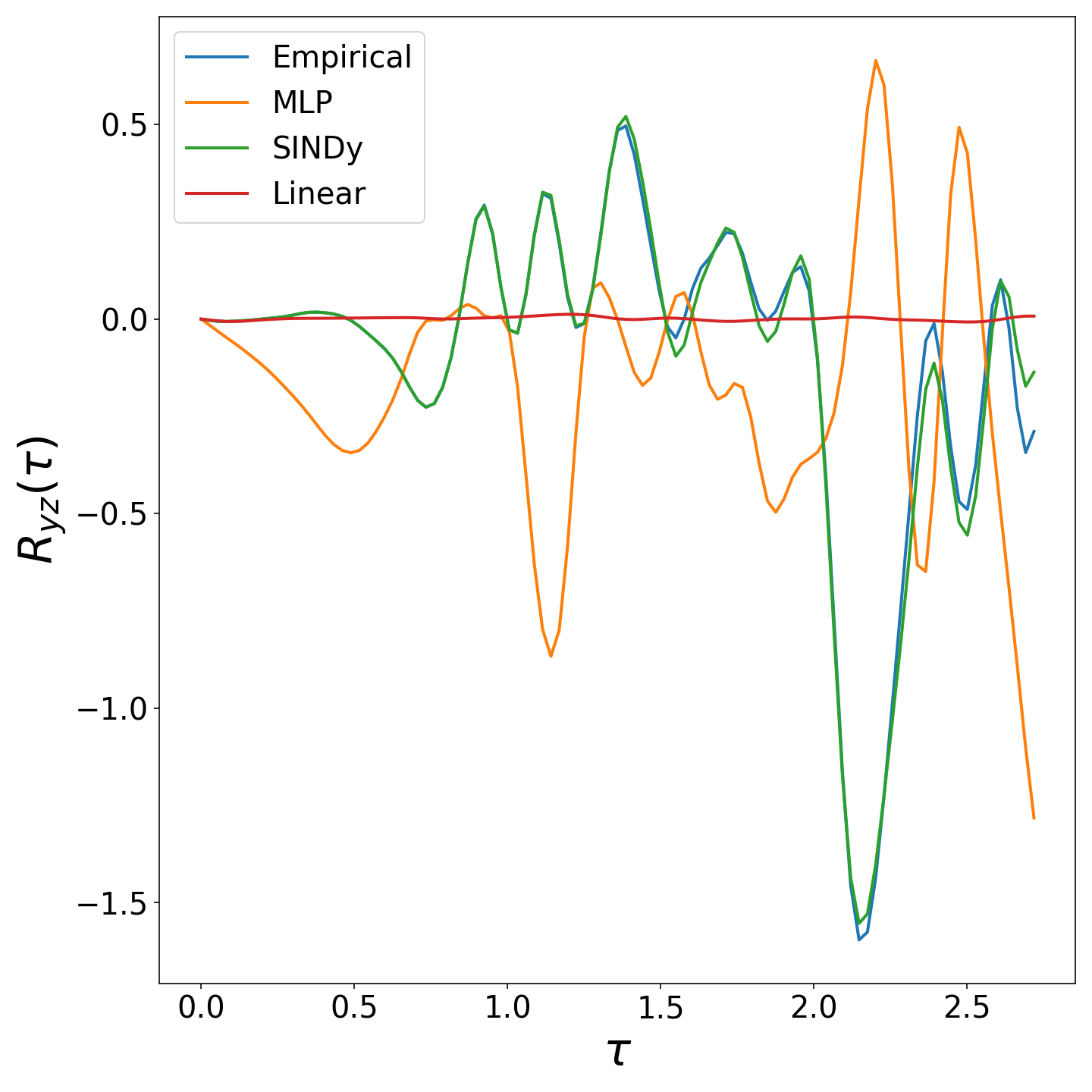}
       \includegraphics[width=0.35\textwidth]{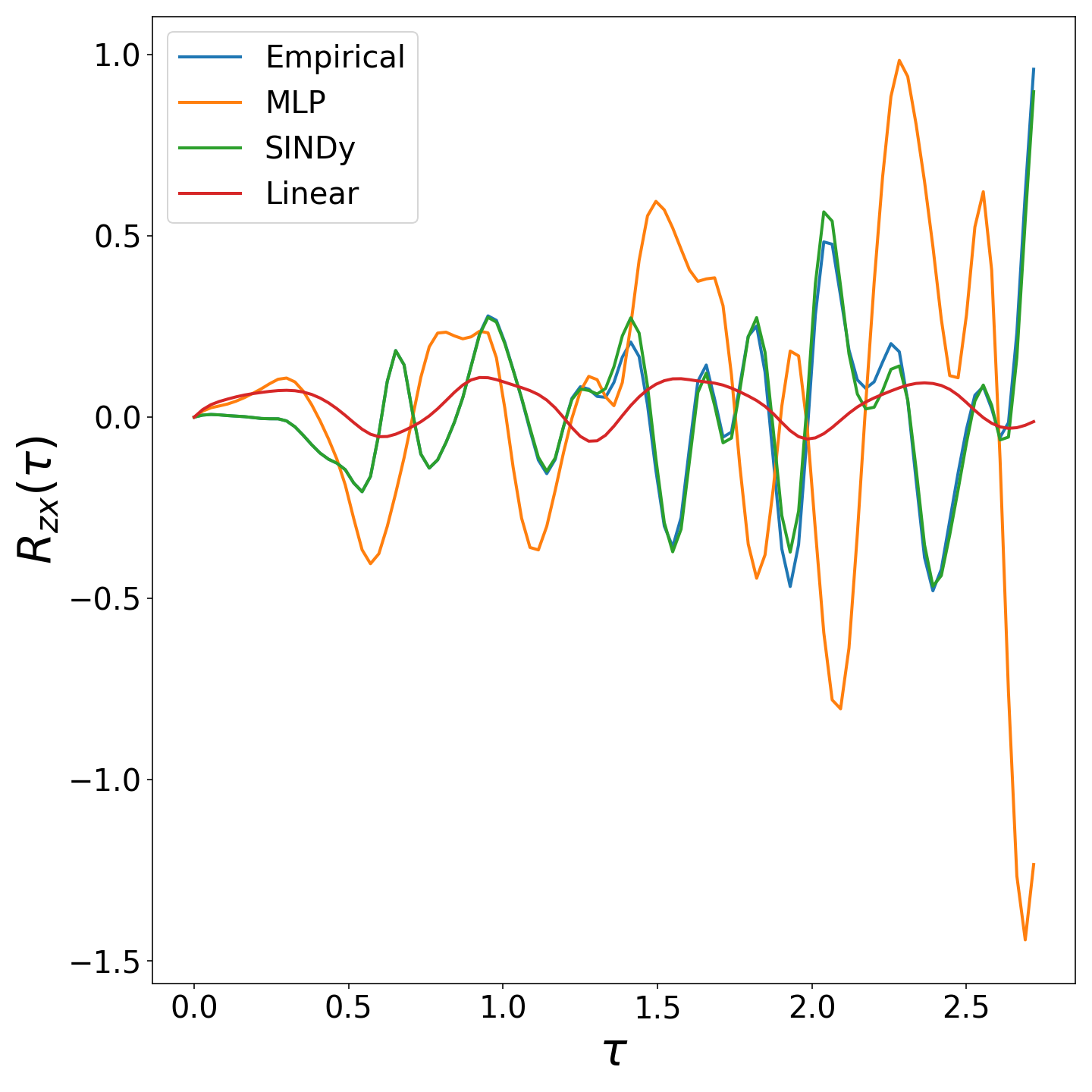}
       \includegraphics[width=0.35\textwidth]{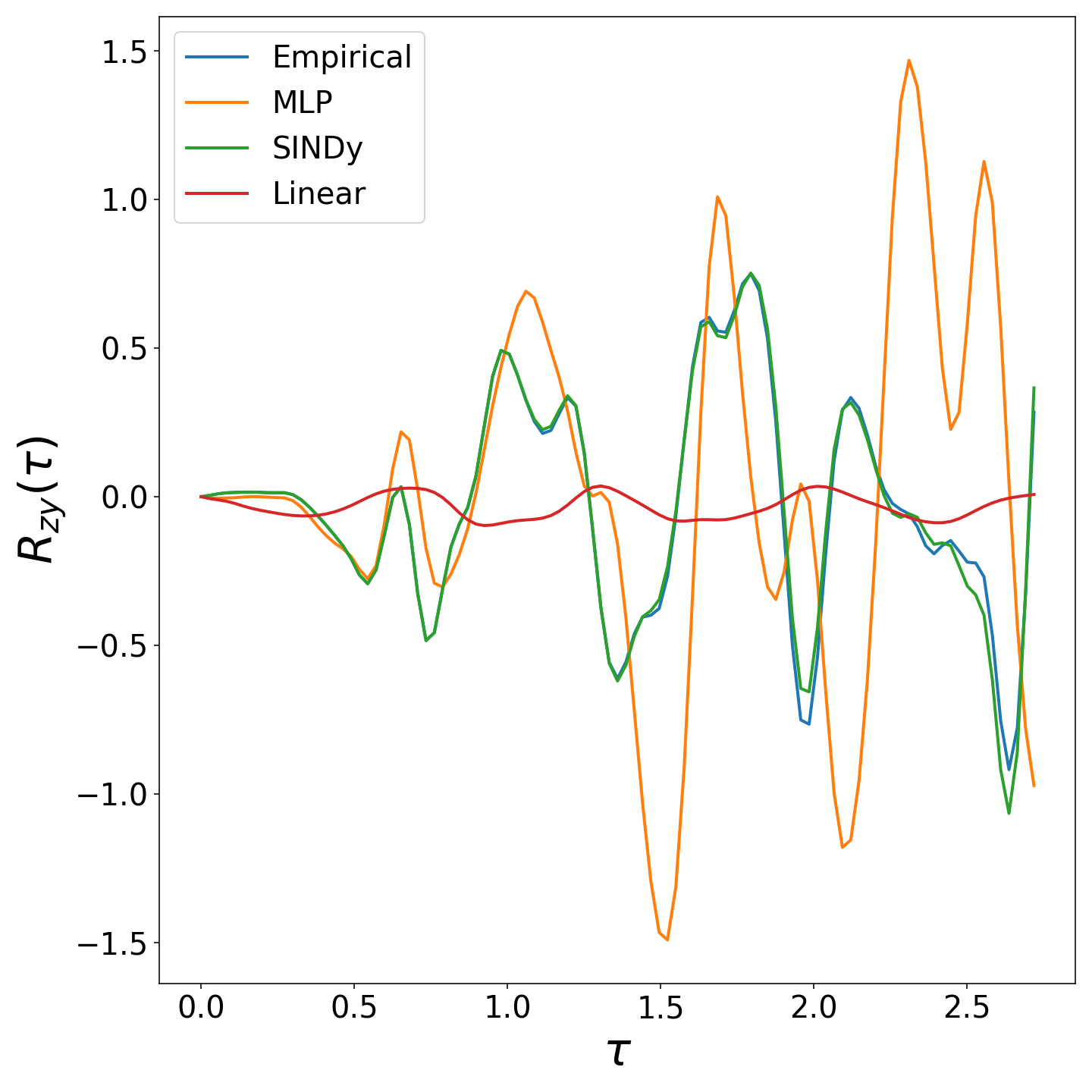}
    \caption{The response functions $\RR_{i\to j}(\tau)$ for $i,j = {x,y,z}$, computed as in the previous figure. We focus here on the non-linear couplings. }
    \label{fig:lor-resp2}
\end{figure}
Looking at the nonlinear couplings tells a slightly different story. SINDy continues to reproduce perfectly the response.
Yet,  while the neural network is able to provide a better picture than the crude linear approximation, at least in terms of shape, the MLP is able to provide at best only a qualitative answer.
It is interesting to underline that in fully nonlinear systems the accurate reconstruction of auto-correlations and of the dimension of the attractor are
not enough to guarantee a good response of the system. 
We have found that some of the cross-correlations are only qualitatively predicted on the long-time by the MLP. That indicates that all the correlations play a non-trivial role in response, in line
with what is found for the weakly non-linear Markov processes.
In Appendix~\ref{app:Lorenz}, we show the cross-correlations.

We want now to build a specific set-up to get more insights on the robustness of the synthetic models.
Let us imagine that the time-series of observations may be affected by systematic errors.
In this case, the trajectories used to compute the response do not correspond to the exact model. We would like to quantify the impact of such an error, notably on the basis of the presence of Chaos.

\begin{figure}[h]
    \centering
                \includegraphics[width=0.35\textwidth]{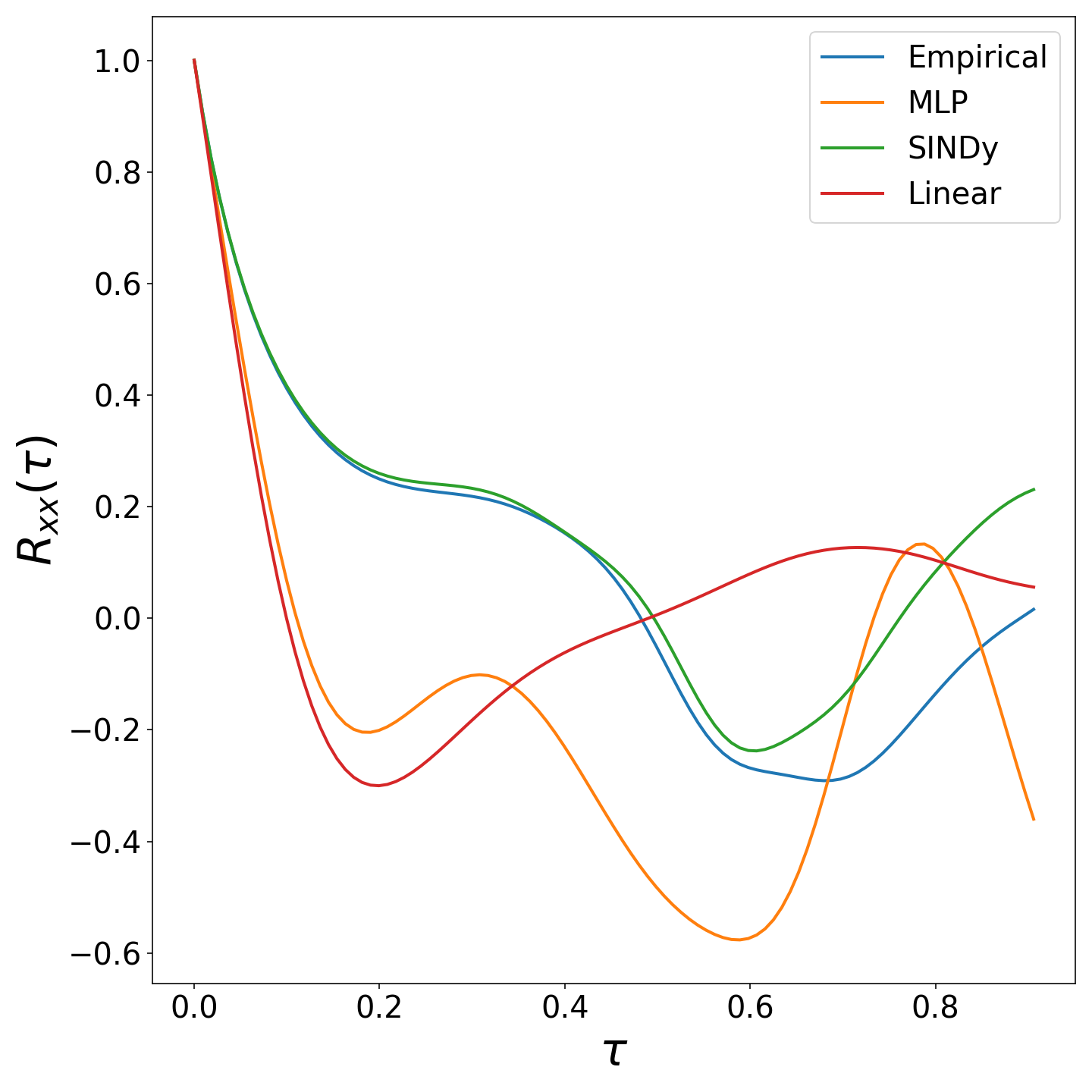}
        \includegraphics[width=0.35\textwidth]{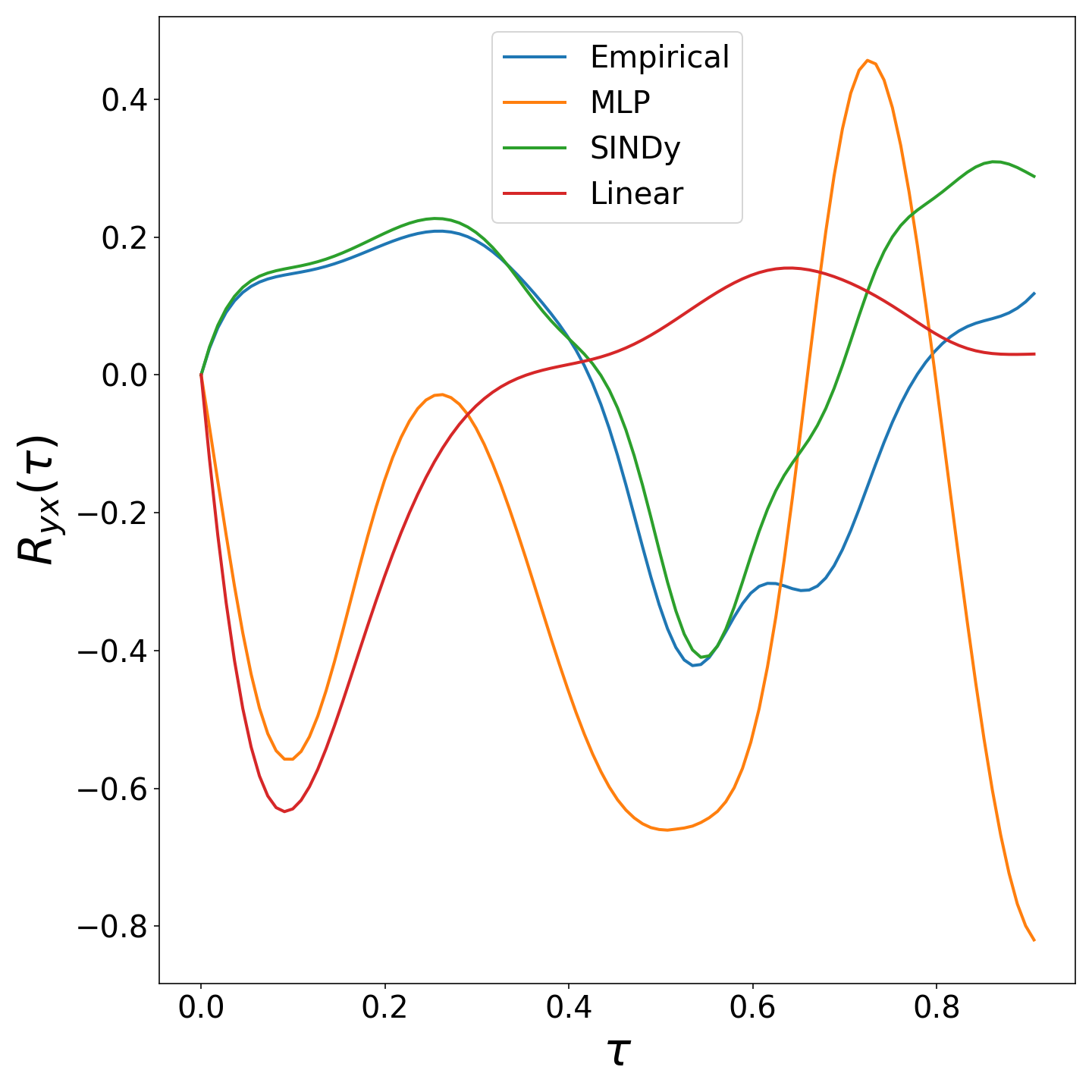}\\
      \includegraphics[width=0.35\textwidth]{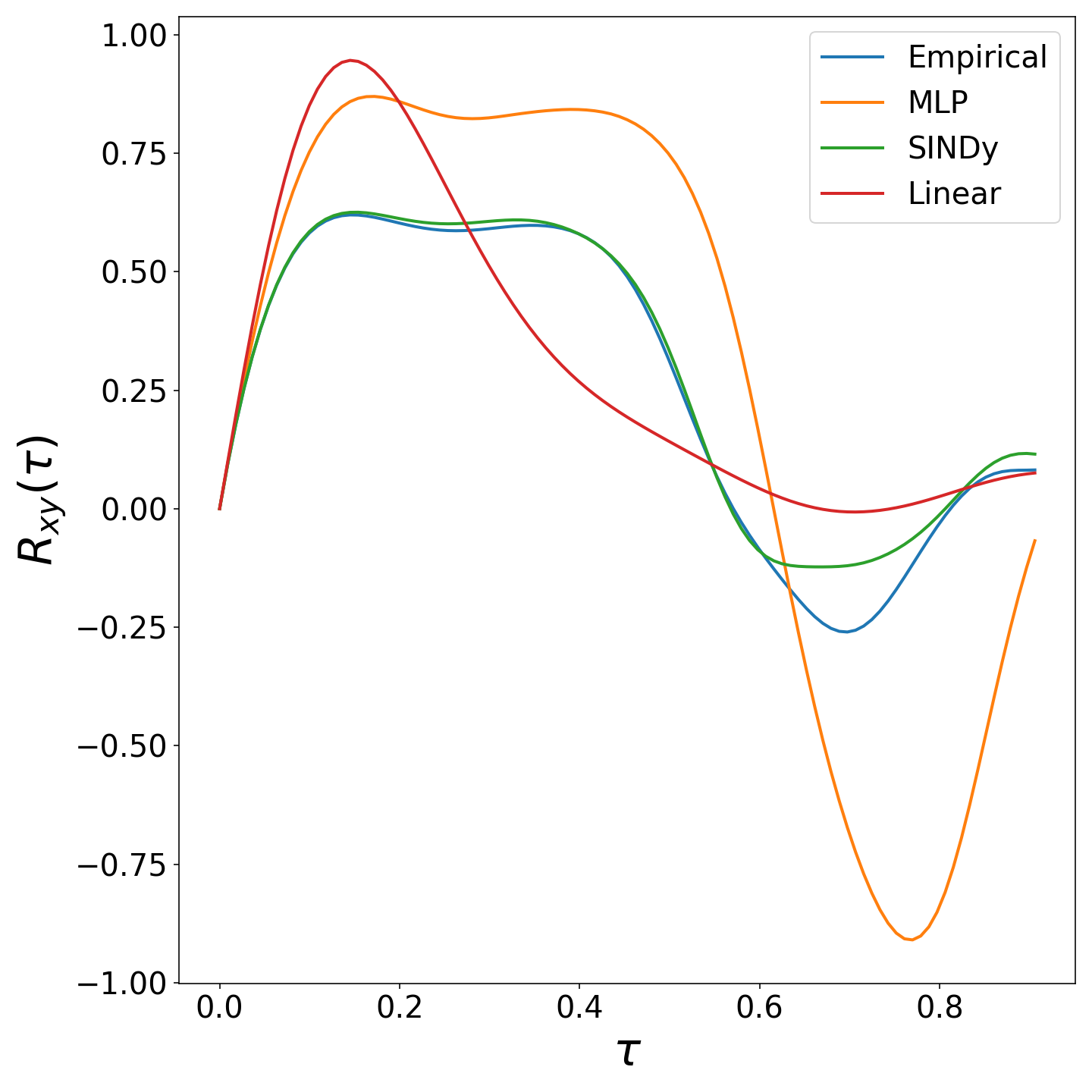}
       \includegraphics[width=0.35\textwidth]{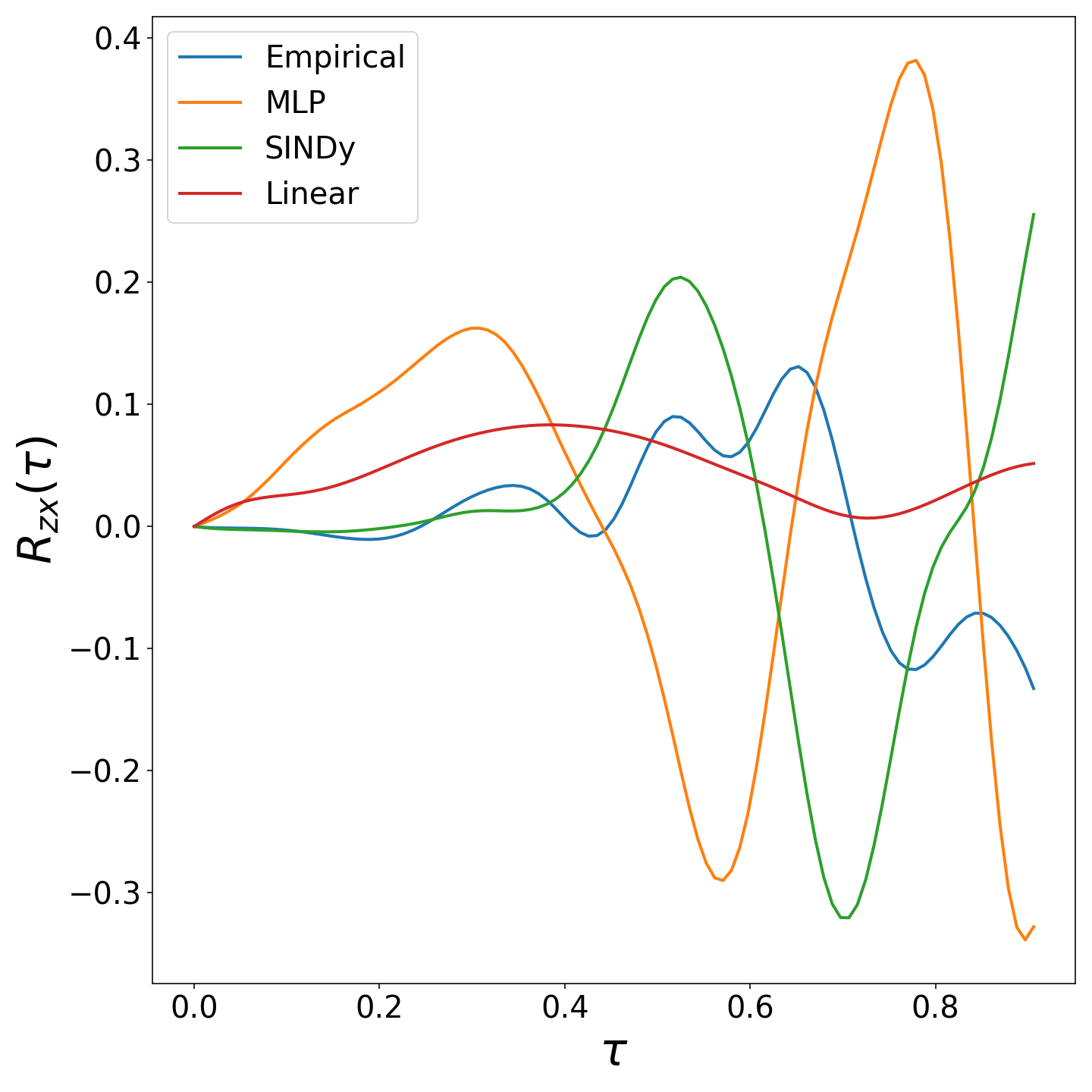}
       \caption{The response functions $\RR_{i\to x}(\tau)$ for $i \in\{x,y,z\}$, computed as in the previous figure.
         We deal with a relatively short time-series here to emphasize the differences. As before, the time-scale is normalized by the Lyapunov
         time $1/ \lambda \approx 500$ units, $\lambda$ is the maximum Lyapunov exponent.  The unperturbed trajectory is computed using the different methods presented.
         The error starts growing at a time of order $\lambda$.}
    \label{fig:lor-resp3}
\end{figure}

To this end, we first build a reference trajectory using a given initial conditions for the MLP and SINDy available models; these trajectories are different from
the trajectory produced by the Lorenz model with same initial conditions, which should be the unbiased observational time-series.
Then we compute the response to a perturbation to those trajectories.
It is worth noting that we have added a small error in the SINDy coefficients, namely at the sixth digit.
In this way, we emphasise the possible differences in the models.
Few examples of the responses obtained in this new set-up are shown in Fig. \ref{fig:lor-resp3}.
The linear approximation has not changed, and we show it just for reference.
The MLP provides results very similar to those shown in Figs. \ref{fig:lor-resp1}-\ref{fig:lor-resp2}.
The MLP-based model has clearly some differences from the true Lorenz system, but it is not sensitive to the way of producing the unperturbed trajectory.
Instead, the small error in the SINDy model is magnified in the present approach. The small error in the coefficients lead to generate a
slightly different unperturbed trajectory and both errors jointly make an impact on the response.
It is clear that in this case SINDy is not able to exactly reproduce the original. In particular, in all cases we notice a departure from the original Lorenz response around $t\sim 500$ which corresponds roughly to the Lyapunov time. It is also interesting to note that in the linear or weakly non-linear couplings, panels (a)-(c), the results obtained by SINDy are not exact but provide a good description of the response. Yet, when considering a fully non-linear coupling as in panel (d), the model is not able anymore to give a good qualitative description of the response on the long-time limit.

It is possible to relate this behaviour to Chaos and the particular expression used to compute the response.
In the present case, we compute the response through the definition (\ref{eq:resp-num}).
In the presence of chaos, the two trajectories ${\bf x}(t)$
and ${\bf x}'(t)$ typically separate exponentially in time, therefore
the mean response is the result of a delicate balance of terms which
grow in time in different directions.  A naive estimate of the  error in the
computation of $\RR_{i\to j}(\tau)$ suggests an increase in time as 
\begin{equation} 
\label{3.16}
e\n(\tau)=\Bigl[ {1 \over N} \sum_{k=1}^N 
\Bigr( { {\delta  x_j(t_k+\tau|t_k)} \over  {\delta  x_i(t_k)}}\Bigl)^2 -
\Bigr( \RR_{i\to j}(\tau)\Bigl)^2 \Bigl]^{1/2}
\sim 
{ {e^{{\lambda_2 \over 2} \tau}} \over  {\sqrt{N}} } \,\, ,
\end{equation}
 where $\lambda_2$ is the generalized Lyapunov exponent of order $2$~\cite{castiglione2008chaos}:
\begin{equation} 
\label{3.17}
\lambda_2= \lim_{\tau \to \infty} {1 \over \tau}
\ln \Biggl \langle \Bigr(
{ {|\delta {\bf x}(\tau)|} \over  {|\delta {\bf x}(0)|} } \Bigl)^2 
\Biggr \rangle \,\, .
\end{equation}
In the above equation  $\delta {\bf x}(\tau)$ is assumed infinitesimal,
i.e. evolving according the linearized dynamics 
(in mathematical terms,  $\delta {\bf x}(\tau)$
is a tangent vector of the system)~\cite{boffetta2002predictability}, and
a lower bound for $\lambda_2$ is given by $2 \lambda$.
Thus very  large $N$ seems to be  necessary,
in order to properly estimate this balance and to compute
$\RR_{i\to j}(\tau)$ for large $\tau$.
This issue is related to an argument made by van Kampen concerning the fluctuation-dissipation theorem~\cite{van1971case,falcioni1990correlation}.

Some remarks are in order.
(i) In the case of linear quasi-Gaussian systems, the response is given in terms of time-correlations, in particular diagonal terms are proportional to auto-correlations. Here we show clearly that in the general nonlinear case a model built from data may reproduce very accurately the auto-correlations, without providing a satisfactory prediction of the response in all cases. 
(ii) Yet, to infer a model from data is possible, provided sufficient data for training are available. If no physical information is used, the model will give a decent representation of causality in terms of response for all linear or weakly nonlinear couplings. 
If one wants to obtain a fully predictive approach, it is needed to combine data and some prior physical modelling  to infer the model. In this case, results are excellent.
(iii) Finally, while it is not too difficult to learn a model able to accurately reproduce the phase-space dynamics, at least for moderately high-dimensional systems, the response turns out to be a very powerful but delicate statistical observable, which may be subject to chaos.

\section{Response theory analysis for high dimensional linear Markov systems}\label{sec:theory}
\begin{figure}[h]
\begin{minipage}{0.7\textwidth}
    \centering\includegraphics[width=\textwidth]{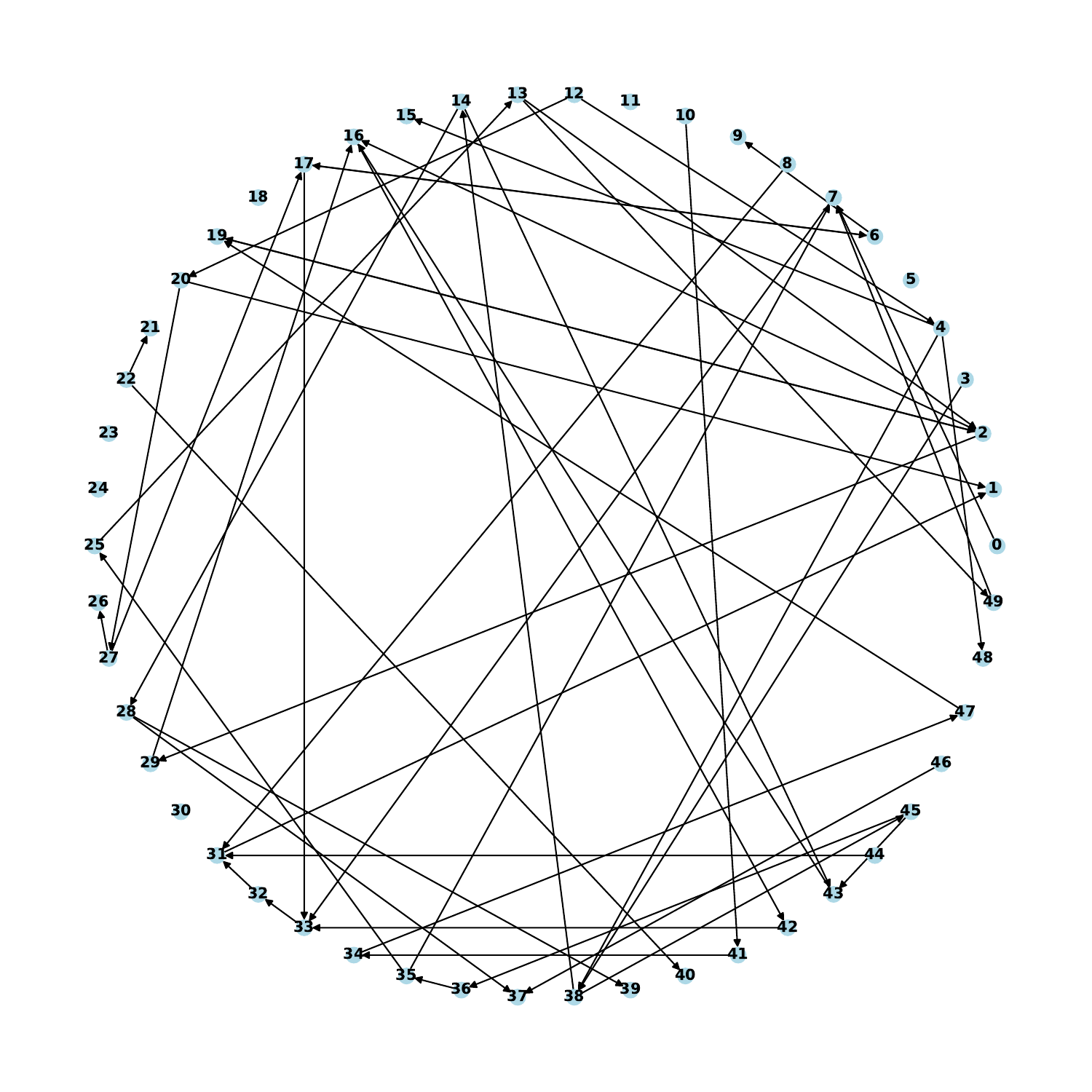}
\end{minipage}
\begin{minipage}{0.29\textwidth}
    \centering\includegraphics[width=\textwidth]{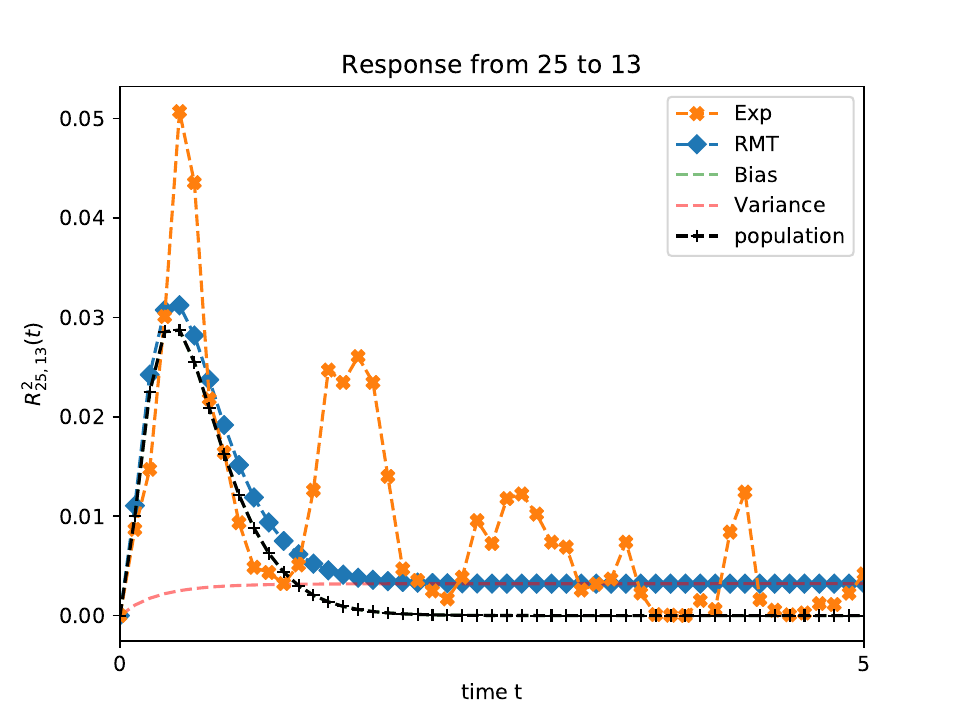}
            \includegraphics[width=\textwidth]{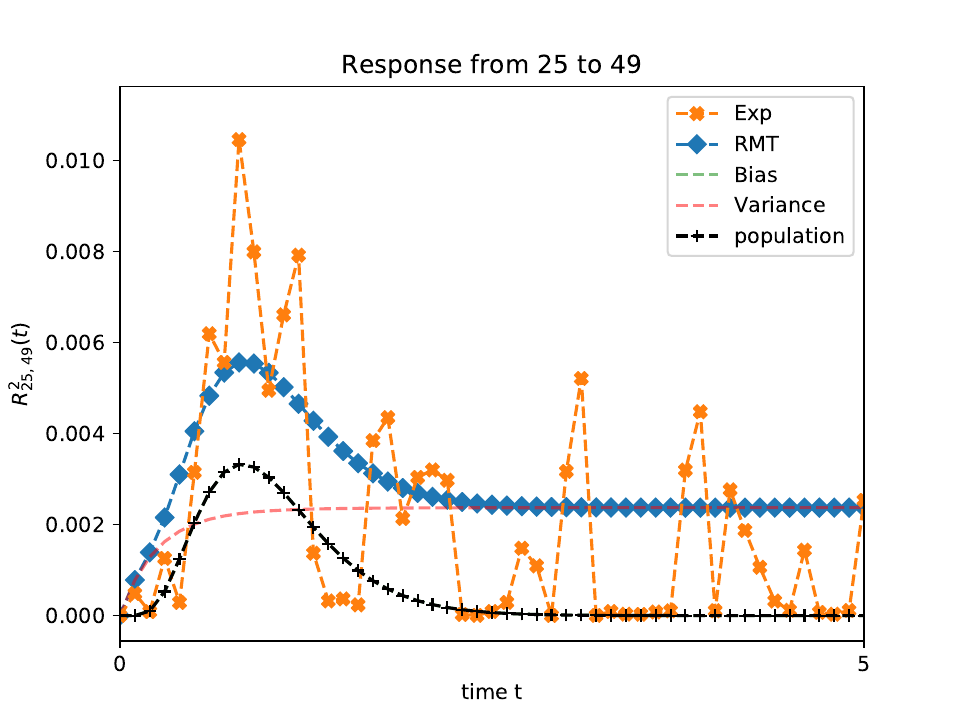}\\
    \centering\includegraphics[width=\textwidth]{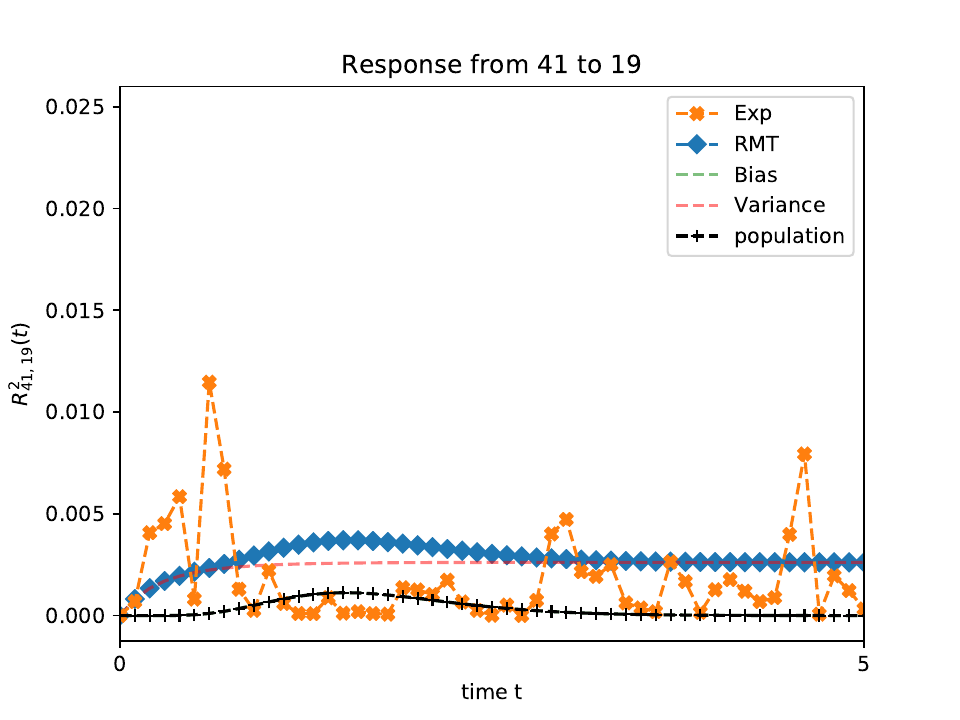}
            \includegraphics[width=\textwidth]{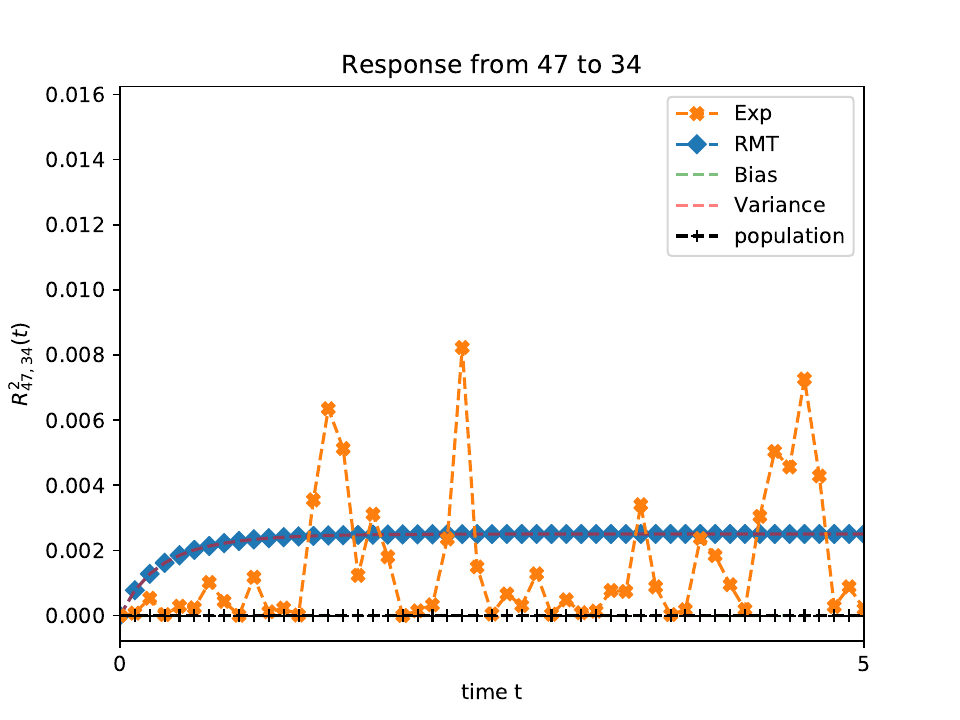}
\end{minipage}
\caption{(left) Random oriented graph of size $d=50$ and output mean connectivity $\bar c = 1.2$. (Right) Comparison between response functions for resp. first, second third NN causal
  links and non-causal links for the last one. The black curves corresponds to the population responses, i.e. in the limit $N\to\infty$ at fixed $d$,
the yellow are the empirical one, while the response predicted by RMT is in blue.}\label{fig:network}
\end{figure}

Up to now, we have considered systems consisting of a few observables with different types of interactions, either linear, weakly non-linear or strongly non-linear leading to chaotic behavior.
In this section we would like to consider the situation where the number of observables in interaction is extensive, and to which extend causal relations among them can be extracted.
This is of interest for instance when facing networks of interactions with unknown structure. Here we will limit ourselves to linear interactions and investigate what the linear response
theory can tell us about causal relations in this context, when the number of observations is proportional to the number of observables.
The goal is to quantify the efficiency of the response theory based causal predictor in an asymptotic regime corresponding
to the case of an extensive number of time series associated to sensors taking place on a large causal graph. 
We assume the underlying process to obey the linear dynamics given by~(\ref{eq:evo_app}) rewritten in a slightly different way in order to account
for some  relaxation of the process explicitly
\begin{equation}\label{eq:dynamics}
\x_{t+1} = \bigl[(1-\epsilon)\I + \epsilon A\bigr]\x_t + \sigma \vect{\eta}_t
\end{equation}
controlled by a damping coefficient $0<\epsilon\ll 1$, while  $A$ is a $d\times d$ incidence matrix, where $A_{ij}=1$ represents an oriented link from $j$ to $i$
and $\sigma$ the standard deviation of the noise, represented here by the
$d$-dimensional normal variable $\eta_{it} = {\mathcal N}(0,1),i=1,\ldots d$, decorrelated both in time and among the coordinates. An exemple of such a network is shown on Figure~\ref{fig:network}.
In order to make this concrete, we can picture this system as an hydraulic network, where only the nodes are observed and behave both as sources and sinks (the noise term),
the oriented flow between nodes being determined by the elements of the incidence matrix, these being not observed.   
The data time series have a length of size $T$ in terms of times steps $t=0,\ldots T$ and we are interested by the
so-called proportional asymptotic limit where both $d,T\to\infty$ with fixed ratio $T/d$. $A$ itself encodes the topology of a sparse random oriented graph of mean connectivity
$\bar c$. The goal is to study causal predictors $\RR(t)$ based on linear response theory.

\subsection{Causal Response functions}
In this case the linear response coincides with the linear regression of $\x_{t+\tau}$ from $\x_t$. For sake of simplicity we will limit the analysis to the ridge regularized regression,
i.e. with $L_2$ penalty. We will refer to this as the linear ridge response (LRR) causal indicator.
Sparse regression techniques, either based on $L1$ or other prior like Gauss-Bernoulli, can be analyzed along similar lines
as the compressed sensing problem~\cite{krzakala2012statistical,ganguli2010statistical} and will be discussed  in a separate work~\cite{ChFu} extending the present formalism
to the sparse causal regression context.  
By convenience, for reasons which will be clearer later on, we denote the ridge penalty
by $\alpha^{-1}$. In order to express the linear response let us introduce some notations. We assume that we are able to extract a set of $N$
independent samples $\{\x_{t(s)},s=1,\ldots N\}$, the proportional scaling limit being characterized by the ratio $\rho \egaldef \frac{N}{d}$. In case we want to use all the available
samples, these cannot then be considered as independent, but
in principle some corrections could be performed on $\rho$, based on the level of correlation between successive samples, so that the following analysis remains valid upon replacing $\rho$ by some
effective one, but we leave aside for sake of simplicity.
We are interested in the empirical estimate of the response function corresponding to the time delay $t$ based on these data.
Let us define the following correlation matrices (assuming $\x_t$ to be centered)
\[
C_t\n \egaldef \frac{1}{N}\sum_{s=1}^N\x_{t(s)+t}\x_{t(s)}^\top
\]
and the resolvent
\[
G\n \egaldef \bigl(\I+\alpha C_0\n\bigr)^{-1},
\]
then the empirical $L_2$ regularized linear response reads
\begin{equation}\label{eq:Rt}
\RR\n(t) = \alpha C_t\n G\n.
\end{equation}
This is an $d\times d$ matrix, and we denote its elements by
\[
\RR_{i\to j}\n(t) \egaldef \e_i^\top \RR\n(t) \e_j
\]
with $\{\e_i,i=1,\ldots d\}$ the canonical vector basis of $\R^d$ and 
where the arrow indicates the causal order.
Based on this we define the squared causal response (SCR) from node $i$ to $j$
\begin{align*}
  \chi_{i\to j}\n(t) &\egaldef \Bigl\langle\vert \e_j^\top \RR_t\n \e_i \vert^2\Bigr\rangle_{\vect{\eta}}  \\[0.2cm]
  &= \alpha^2\Bigl\langle\Tr\Bigl[G\n\e_i\e_i^\top G\n {C_t\n}^\top\e_j\e_j^\top C_t\n\Bigr]\Bigr\rangle_{\vect{\eta}}
\end{align*}
where $\langle\rangle_{\vect{\eta}}$ denotes an averaging over the noise.
We denote by $\chi_{i\to j}(t)$ its corresponding asymptotic limit in the proportional scaling.
Given some non-negative threshold $h$ and some inverse temperature $\beta$, we are interested by the LRR causal predictor which we define as
\begin{equation}\label{eq:causal_pred}
p_{i\to j}(t) \egaldef \sigma\Bigl[\beta(\chi_{i\to j}(t)-h)\Bigr],
\end{equation}
where $\sigma(x) \egaldef 1/(1+e^{-x})$ is the sigmoid function.
In order to obtain this limit we first need the covariance of $\x_t$ at coinciding time in the population limit defined as
\[
C \egaldef \lim_{N\to\infty} C_0\n.
\]
From the definition~(\ref{eq:dynamics}) of the underlying process $C$ satisfies a discrete-time Lyapounov equation
\[
C = MCM^\top +\sigma^2\I = \sigma^2\sum_{\tau=0}^\infty M^\tau {M^\tau}^\top
\]
where $M = (1-\epsilon)\I+\epsilon A$. We assume that the spectral radius of $M$ is smaller than $1$ for $C$ to be well defined.
In addition we have
\begin{equation}\label{eq:C_tn}
C_t\n = M^t C_0\n + \frac{\sigma}{N}\sum_{s=1}^N \sum_{\tau=0}^{t-1}M^{t-\tau-1}\vect{\eta}_{t(s)+\tau}\ \x_{t(s)}^\top.
\end{equation}
Using this we obtain for the SCR 
\[
\chi_{i\to j}\n(t) = \alpha^2\Tr\Bigl[G\n\e_i\e_i^\top G\n \Bigl({C_0\n}^\top {M^t}^\top\e_j\e_j^\top M^tC_0\n 
    +\frac{\sigma^2}{N}\sum_{\tau=0}^{t-1}\e_j^\top M^{t-\tau-1}{M^{t-\tau-1}}^\top\e_j C_0\n\Bigr)\Bigr]
\]
Thanks to the identity $\alpha G\n C_0\n = \I-G\n $ we finally obtain an expression which involve only the resolvent:
\begin{align}
\chi_{i\to j}\n(t) &= \Tr\Bigl[(\I-G\n)\e_i\e_i^\top (\I-G\n){M^t}^\top\e_j\e_j^\top M^t\Bigr] \nonumber\\[0.2cm]
  &+\frac{\alpha\sigma^2}{N} \Tr\Bigl[G\n(\I-G\n)\e_i\e_i^\top\Bigr]\times\sum_{\tau=0}^{t-1}
  \e_j^\top M^{t-\tau-1}{M^{t-\tau-1}}^\top\e_j \label{eq:cr}
\end{align}
where the first term which is ${\mathcal O}(1)$ corresponds to a bias independent of the noise while the second one which is ${\mathcal O}(t/N)$ is a variance term resulting from
averaging over the noise. Hence we expect the second term to become meaningful for a regime where $t/N$ is $\mathcal{O}(1)$, since 
averaging over the noise leads us to get rid of stochastic contributions which are typically of order $1/\sqrt{N}$. 
Note, that the average of the noise is in principle problematic, since it appears explicitly in~(\ref{eq:C_tn}) but is also present implicitly in the $\x_{t(s)}$
i.e. in $C_0\n$ leading to additional correlations that we do not take into account. The reason these are actually negligible relies actually on the hypothesis that
$t(s+1)-t(s)$ is sufficiently large to insure independence between $\x_{t(s)}$ and $\x_{t(s+1)}$.

\subsection{Asymptotic efficiency of LRR causal predictor}
In order to characterize the efficiency of the LRR causal predictor we need to introduce additional quantities. 
Assuming that the fluctuations of the linear response are Gaussian we need to characterize its bias and variance,
conditionally to whether the causal link from $i\to j$ exists or not. Let us define by $\f_{\to i}(t)$ the vector of \emph{true} underlying response coefficients associated to output node $i$
and time delay $t$, corresponding to what would be observed in absence of noise
\[
\f_{\to i}(t) \egaldef \e_i^\top M^t.
\]
Considering $\{\RR_{j\to i}\n(t),j=1,\ldots d\}$ as a set of independent random variables,
we assume these to be Gaussian conditionally to $f_{j\to i}(t)$, i.e. of the form
\[
\RR_{j\to i}\n(t) = \mu_{j\to i}\n(t) + \sigma_{\to i}\n(t)z_{j\to i}
\]
where $z_{j\to i}\sim\N(0,1)$ are iid. $\mu_{j\to i}\n(t)$  and ${\sigma_{\to i}\n}^2(t)$ are respectively the bias and variance of $\RR_{j\to i}\n(t)$ given by 
\begin{align*}
  \mu_{j\to i}\n(t) &=   \frac{f_{j\to i}(t)}{\Vert\f_{\to i}(t)\Vert^2}\Bigl\langle \f_{\to i}^\top(t)\RR_{\to i}\n(t) \Bigr\rangle_{\vect{\eta}}, \\[0.2cm]
  {\sigma_{\to i}\n}^2(t) &= \Bigl\langle\Vert\RR_{\to i}\n(t)\Vert^2\Vert\Bigr\rangle_{\vect{\eta}} -
  \frac{\Bigl\langle\f_{\to i}^\top(t)\RR_{\to i}\n(t)\Bigr\rangle_{\vect{\eta}}^2}{\Vert\f_{\to i}(t)\Vert^2}.
\end{align*}
$f_{j\to i} = \e_i^\top M^t\e_j$ is considered to be given but unknown and the other quantities like the overlap between $\f_{\to i}$ and $\RR_{\to i}\n(t)$ 
take simple form in term of the resolvent:
\begin{align}
  \Vert\f_{\to i}(t)\Vert^2 &= \e_i^\top M^t {M^t}^\top\e_i,\label{eq:signal}  \\[0.2cm]
\Bigl\langle \f_{\to i}^\top(t)\RR_{\to i}\n(t) \Bigr\rangle_{\vect{\eta}} &= \Tr\bigl[(\I-G\n){M^t}^\top \e_i\e_i^\top M^t\bigr],\label{eq:overlap}\\[0.2cm]
  \Bigl\langle \Vert\RR_{\to i}\n(t)\Vert^2\Bigl\rangle_{\vect{\eta}} &= \Tr\Bigl[(\I-G\n)^2 {M^t}^\top\e_i\e_i^\top M^t\Bigr]
  +\frac{\alpha\sigma^2}{N} \Tr\Bigl[G\n(\I-G\n)\Bigr]\times\sum_{\tau=0}^{t-1}\e_i^\top M^{t-\tau-1}{M^{t-\tau-1}}^\top\e_i.\label{eq:Respnorm}
\end{align}  
As illustrated on Figure~\ref{fig:network}, each coefficient $\RR_{i\to j}$ maintains its random nature in the asymptotic limit, such that  the asymptotic estimation~(\ref{eq:cr_rmt}) of the 
squared response averaged over the noise cannot be leveraged directly to analyze the asymptotic behavior of the LRR
causal predictor. Instead, both $\mu_{j\to i}$ and $\sigma_{\to i}^2$ become deterministic in the asymptotic regime and can be
fully determined analytically thanks to standard random matrix formulas~\cite{MP_law,ledoit2011eigenvectors} upon solving RMT equations as stated in Appendix~\ref{App:RMT}.  
Then the performance of the causal predictor~(\ref{eq:causal_pred}) being a simple function of these bias and variance is fully characterized.
\begin{figure}[h]
    \centering\includegraphics[width=0.49\textwidth]{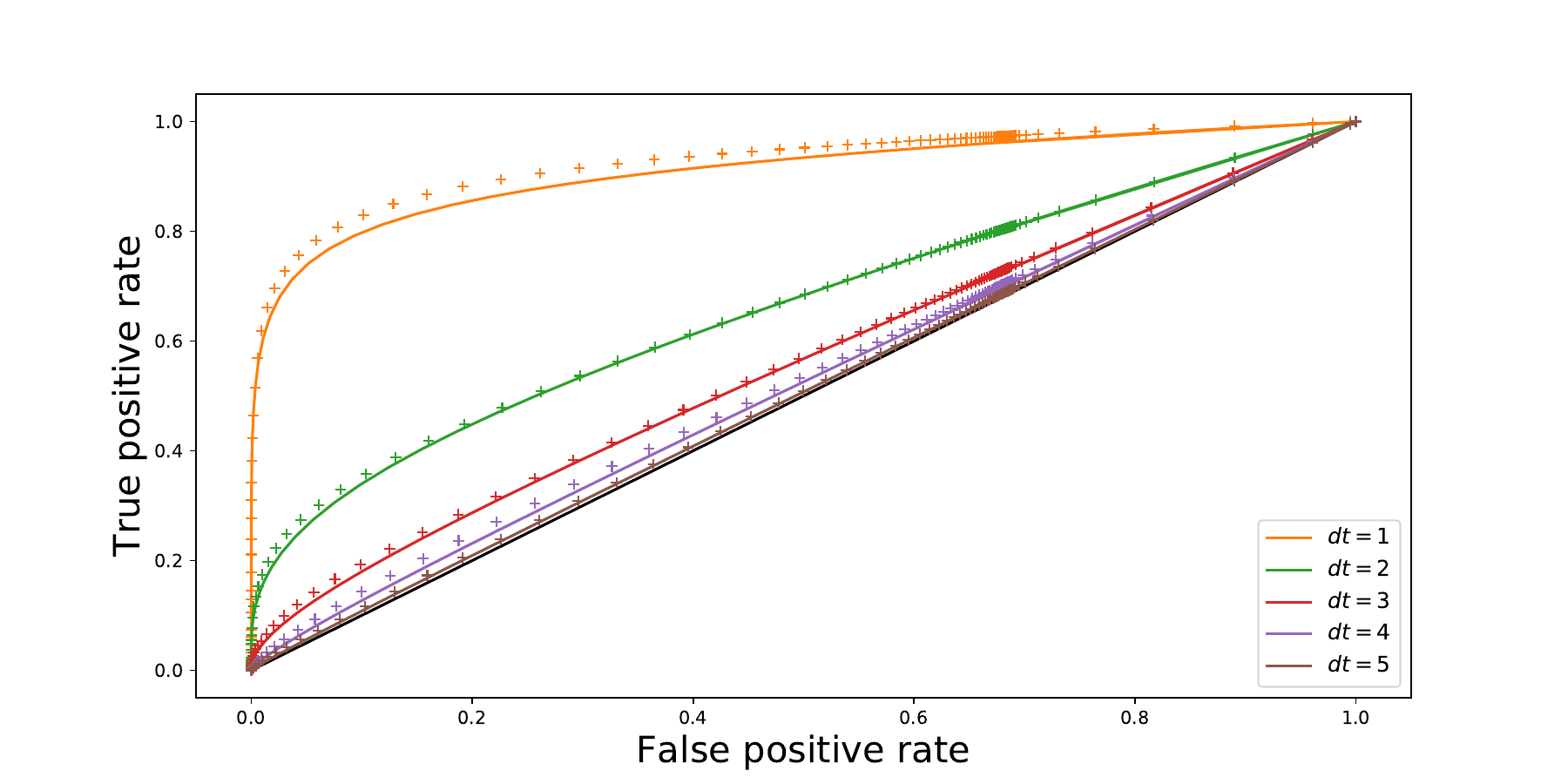}
    \includegraphics[width=0.49\textwidth]{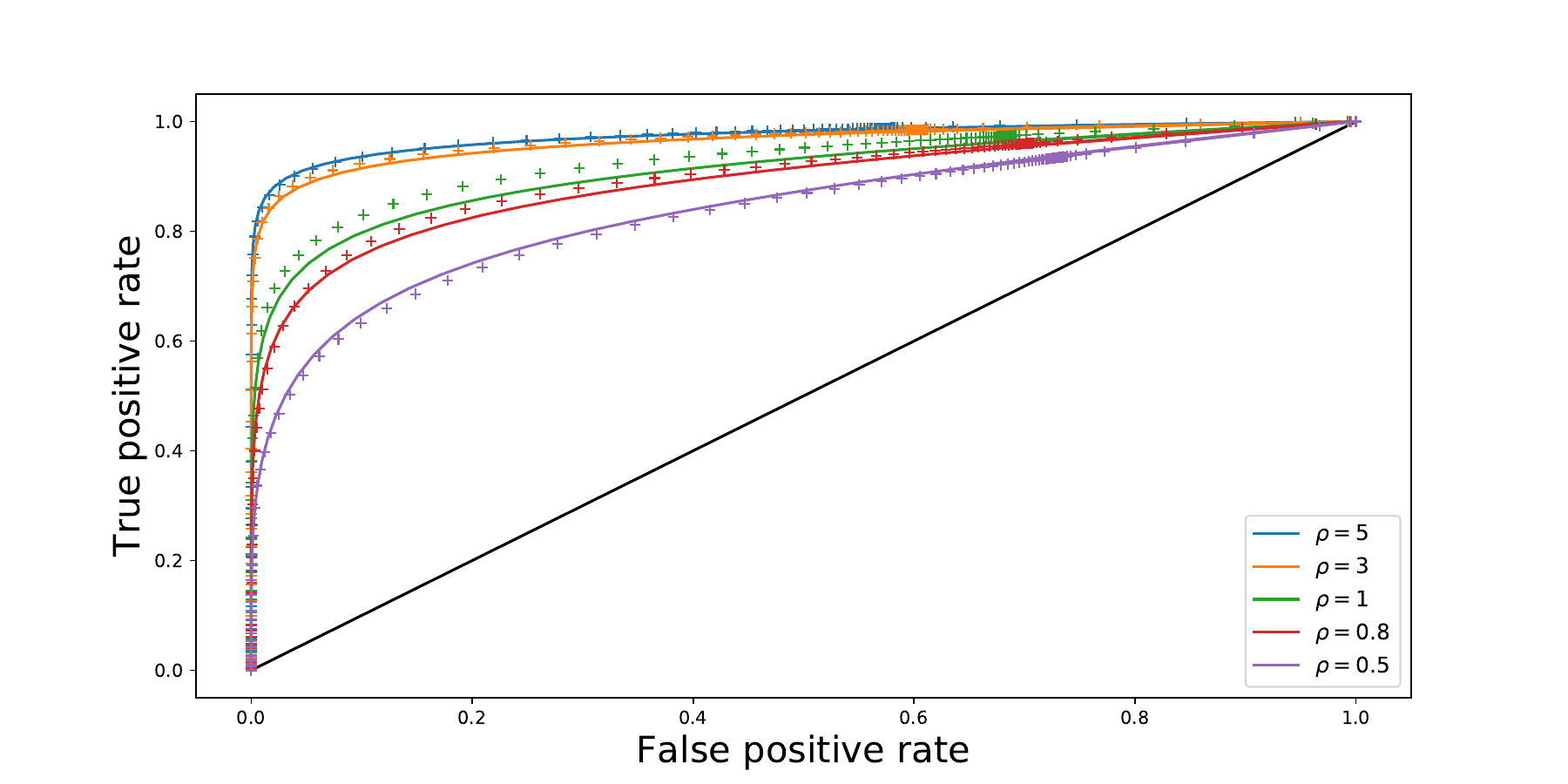}
    \caption{\label{fig:roc} ROC curve  of the LRR causal predictor
      obtained for a network of size $d=100$ mean connectivity $\bar c = 1.7$, inverse ridge penalty $\alpha=10$,
      when varying respectively the time delay for the response on the left with $\rho = 2.5$ and the aspect ratio $\rho$ with $dt=1$ on the right panel.}
\end{figure}
Equation~(\ref{eq:overlap2}) given in Appendix~\ref{App:RMT}
along with~(\ref{eq:signal}) fully determine $\mu_{\to i}(t)$ and $\sigma_{\to i}(t)$.
In turn partitioning the set of links $\{i,j\}(t) = \bar\IC(t)+\IC(t)$ corresponding respectively to the set of zero and non-zero entries in $M^t$ 
leads to the following expressions for the expectation of true positive and false positive rates of causal links given by the LRR causal predictor:
\begin{align*}
  \overline{TP}(h,t) &= \Bigl\langle{\mathbb E}_{\eta\sim\N(0,1)}\Bigl[\sigma\Bigl(\beta\bigl(\vert\mu_{i\to j}(t)+\sigma_{\to j}(t)\eta\vert^2\bigr) - h\Bigr)\Bigl]\Big\rangle_{(i,j)\in\IC(t)} \\[0.2cm]
  \overline{FP}(h,t) &= \Bigl\langle{\mathbb E}_{\eta\sim\N(0,1)}\Bigl[\sigma\Bigl(\beta\bigl(\vert\mu_{i\to j}(t)+\sigma_{\to j}(t)\eta\vert^2\bigr) - h\Bigr)\Bigl]\Big\rangle_{(i,j)\in\bar\IC(t)}
\end{align*}
where $h$ is an arbitrary threshold, and the empirical average over edges, i.e. implicitly on $f_{i\to j}(t)$ is denoted by $\langle\cdot\rangle$.
A standard way to characterize the behavior of this causal predictor is to display the Receiver Operating Characteristic (ROC) curve,
i.e. by plotting number of true positive as a function of number of false positive when varying $h$ in~(\ref{eq:causal_pred}). The comparison between experiments and
asymptotic results is shown on Figure~\ref{fig:roc}, denoting a rapid onset of the asymptotic regime despite the high sparsity of the signal for small time response
functions while for larger time the noise dominates rapidly rendering the LR causal predictor inefficient.

\section{Conclusions}
In this work, we have investigated in some generic examples to which extent and with what accuracy the response theory combined with ML techniques can be used to infer  
causal links between different components of a physical system from data.
The approach is grounded in the statistical physics framework, which allows us to use the
linear response of the system as a proxy for doing interventional causality experiments.
This can be done both for linear and nonlinear systems, provided that we work with a functional representation of the dynamical system in the latter case. 

For linear systems, conceptually the problem is solved in terms of the covariance matrix, which fully determined the response function.
As expected, we have shown that in this case, for linear stochastic Markov processes, it is possible to efficiently extract the response from data with different techniques.
The use of a sparse regularized neural network turned out to be particularly efficient, with a very small number of parameters needed.

In the nonlinear case, the response cannot be calculated from the covariance, but in principle we should access to the whole invariant measure.
We have studied both a stochastic system perturbed by a nonlinear potential, and the Lorenz63 system, paradigm of the fully chaotic systems.
Although both systems are nonlinear, they exhibit different behavior regarding causal response identification. There are clear distinctions between these two systems to explain that:
the first system is stochastic, whereas the second is deterministic and  in the first system, the non-linearity does not couple the variables—unlike
in the Lorenz63 model, where such coupling occurs. In addition the
nonlinear Markov system possesses an underlying linear structure, which makes it possible to capture the system’s response in a qualitative sense, even through a linear approximation.
In this setting, various data-driven approaches prove effective.
On one hand, if one can physically identify a suitable functional basis for the system’s state vector, the response can be
reconstructed with near-exact accuracy. On the other hand, in a purely data-driven framework, powerful recurrent architectures—such as reservoir computing—can still provide a reasonably good agreement.

The situation is  different in the chaotic case. Here, the linear approximation fails entirely and offers no insight into the actual response or causality. A properly parameterized neural
network can successfully reconstruct the phase dynamics in detail, but it still only delivers a qualitative account of the full response.
By contrast, sparse identification makes it possible to infer the correct model directly from data, starting from a large dictionary of candidate functions, as long as the relevant
terms are included. The resulting model is nearly indistinguishable from the true one.
As seen in these experiments, the response of a system is a particularly delicate statistical observable—far more challenging to characterize accurately
than other two-point statistics such as auto-correlations.
This calls for special care when analyzing responses in complex systems. Nevertheless, we have shown that a physics-guided approach can indeed yield highly accurate results.

Finally, we have also considered the case in which the system is simple enough (linear) but the number of degrees of freedom is so high that it imposes a lack of observability.
In this case, it is possible in principle to access to the response via covariance, but we have not enough data to compute it accurately. That may be relevant for network applications.
Leveraging the asymptotic limit we have dealt theoretically with this case in the Random Matrix theory framework. We have shown that it is possible to obtain a rather accurate description of
the response in most of the cases, adding some form of regularization. 

Let us finally conclude with some perspectives. First we could explore the  relation, if any, between this physics based approach and 
existing ones developed in ML regarding time series~\cite{assaad2022survey} or make systematic comparison of its efficiency on standard benchmark data. Dictionary
or RNN based approaches developed in Section~\ref{sec:chaotic_systems} designed to cope with the non-linear dynamical cases point toward using
methods which could presumably have something to do with the HSIC criterion~\cite{gretton2005},
which requires further investigations. Concerning the theoretical setting developed in Section.~\ref{sec:theory} we will as already mentioned extend it to sparse
regularization settings in a separate work. Finally we expect these methods to be relevant for applications and consider to test them in particular on geophysics problems.

\section{Acknowledgements}
The project has received the funding by French ANR grant SPEED  (ANR-20-CE23-0025-01) and
the ANR grant SCALP  (ANR-24-CE23-1320).

\bibliographystyle{unsrt}
\bibliography{biblio}
\appendix


\section{Response theory}
\label{app:resp-teo}
We report a derivation of a general FDR,  based on previous works~\cite{marconi2008fluctuation,baldovin2022extracting}  Consider a dynamical system $ {\bf x}(0) \to {\bf x}(t)=S^t
{\bf x}(0)$ with states ${\bf x}$ belonging to a $N$-dimensional
vector space.  We consider the possibility
that the time evolution  could be described by stochastic differential equations.  We
assume that the system is mixing and that the invariant probability
distribution $\rho({\bf x})$ enjoys some regularity property, while no assumption is made on $N$.
The purpose is to express the average response of a generic observable $A$
to a perturbation, in terms of suitable correlation functions,
computed according to the invariant measure of the unperturbed system.
The first step is to study the behavior of a single component of ${\bf  x}$,
 say $x_i$, when the system, described by $\rho({\bf x})$, is
subjected to an initial (non-random) perturbation 
${\bf  x}(0) \to {\bf x}(0) + \Delta {\bf x}_{0}$~\footnote{The study of an
  ``impulsive'' perturbation is not a severe limitation, for instance in
  the linear regime the (differential) linear response describes the effect of a generic perturbation.}.  This
instantaneous kick modifies the density of the system into
$\rho'({\bf x})$, which is related to the invariant
distribution by $\rho' ({\bf x}) = \rho ({\bf x} - \Delta {\bf x}_0)$.
We introduce the probability of transition from ${\bf x}_0$ at time
$0$ to ${\bf x}$ at time $t$, $W ({\bf x}_0,0 \to {\bf x},t)$. For a
deterministic system, with evolution law $ {\bf x}(t)=S^{t}{\bf
  x}(0)$, the probability of transition reduces to $W ({\bf x}_0,0 \to
{\bf x},t)=\delta({\bf x}-S^{t}{\bf x}_{0})$, where $\delta(\cdot)$ is the
Dirac delta function.  Then we can write an expression for the mean value of
the variable $x_i$, computed with the density of the perturbed system:
\begin{equation}
\label{3.4}
\Bigl \langle x_i(t) \Bigr \rangle ' = 
\int\!\int x_i \rho' ({\bf x}_0) 
W ({\bf x}_0,0 \to {\bf x},t) \, d{\bf x} \, d{\bf x}_0  \; .
\end{equation}
The mean value of $x_i$ during the unperturbed evolution can be written in
a similar way:
\begin{equation}
\label{3.5}
\Bigl \langle x_i(t) \Bigr \rangle = 
\int\!\int x_i \rho ({\bf x}_0) 
W ({\bf x}_0,0 \to {\bf x},t) \, d{\bf x} \, d{\bf x}_0  \; .
\end{equation}
Therefore, defining $\overline{\delta x_i} =  \langle x_i \rangle' -
\langle x_i \rangle$, we have:
\begin{eqnarray}
\label{3.6}
\overline{\delta x_i} \, (t)  &=&
\int  \int x_i \;
\frac{\rho ({\bf x}_0 - \Delta {\bf x}_0) - \rho ({\bf x}_0)}
{\rho ({\bf x}_0) } \;
\rho ({\bf x}_0) W ({\bf x_0},0 \to {\bf x},t) 
\, d{\bf x} \, d{\bf x}_0 \nonumber \\
&=& \Bigl \langle x_i(t) \;  F({\bf x}_0,\Delta {\bf x}_0) \Bigr \rangle
\end{eqnarray}
where
\begin{equation}
\label{3.7}
F({\bf x}_0,\Delta {\bf x}_0) =
\left[ \frac{\rho ({\bf x}_0 - \Delta {\bf x}_0) - \rho ({\bf x}_0)}
{\rho ({\bf x}_0)} \right] \; .
\end{equation}
Note that the system  is assumed to be mixing, so that the decay to zero of the
time-correlation functions prevents any departure from equilibrium.
 
For an infinitesimal perturbation $\delta {\bf x}(0) = (\delta x_1(0)
\cdots \delta x_N(0))$, the function in (\ref{3.7}) can be expanded to
first order, if $\rho({\bf x})$ is non-vanishing and differentiable,
and one obtains:
\begin{eqnarray}
\label{3.8}
\overline{\delta x_i} \, (t)  &=&
- \sum_j \Biggl
\langle x_i(t) \left. \frac{\partial \ln \rho({\bf x})}{\partial x_j} 
\right|_{t=0}  \Biggr \rangle \delta x_j(0) \nonumber \\
&\equiv&
\sum_j \RR_{j\to i}(t) \delta x_j(0)
\end{eqnarray}
which defines the linear response 
\begin{equation}
\label{3.9}
\RR_{j\to i}(t) = - \Biggl \langle x_i(t) \left.
 \frac{\partial \ln \rho({\bf x})} {\partial x_j} \right|_{t=0}
\Biggr  \rangle 
\end{equation} 
of the variable $x_i$ with respect to a perturbation of $x_j$.
One can easily repeat the computation for a generic observable
$A({\bf x})$, obtaining:
\begin{equation}
\label{3.10}
\overline{\delta A} \, (t)  = -\sum_j
\Biggl \langle A({\bf x}(t)) \left.
\frac{\partial \ln \rho({\bf x})} {\partial x_j} \right|_{t=0} 
\Biggr \rangle \delta x_j(0) \,\, .
\end{equation}

In the above derivation of the FDR, we did not use any approximation
on the evolution of $\delta {\bf x}(t)$.  Starting with the exact
expression (\ref{3.6}) for the response, only a linearization of the
initial perturbed density is needed, and this implies only the
smallness of the initial perturbation.  From the evolution of the
trajectories difference, one can  define the maximum Lyapunov
exponent $\lambda$, by considering  the positive quantities $|\delta {\bf
x}(t)|$, so that at small $|\delta {\bf x}(0)|$ and large enough $t$
one can write
\begin{equation}
\label{3.11}
\Bigl \langle \ln |\delta {\bf x}(t)|\Bigr \rangle \simeq 
\ln |\delta {\bf x}(0)| + \lambda t \,\, .
\end{equation}
Differently, in the derivation of the FDR, one deals with averages of quantities
with sign, such as $\overline{\delta {\bf x}(t)}$.  This apparently
marginal difference is very important and underlies the possibility of deriving
 the FDR without incurring in  van Kampen's objection~\cite{falcioni1990correlation,van1971case}.
 
At this point one could object that in  chaotic deterministic
dissipative systems the above machinery cannot be applied, because the
invariant measure is not smooth. 
It is worth emphasising that, even though
in chaotic dissipative systems the invariant
measure is singular~\cite{paladin1987anomalous},  the previous derivation of the FDR is
still valid if one considers perturbations along the expanding
directions. 
In addition, one is often interested in some specific
variables, so that a projection is performed, making irrelevant the
singular character of the invariant measure~\cite{ruelle1998general}.
In these
cases, a general response function has two contributions,
corresponding respectively to the expanding (unstable) and the
contracting (stable) directions of the dynamics. The first
contribution can be associated to some correlation function of the
dynamics on the attractor (i.e. the unperturbed system).
Nevertheless, a small amount of noise, always present in physical systems and numerical simulations, smoothens the $\rho({\bf x})$
and the FDR can be derived.  
Then, 
the assumption on the smoothness of
the invariant measure along the unstable manifold still allows to
avoid subtle technical difficulties~\cite{ruelle1998general}.

\section{Lorenz system Cross-correlations}\label{app:Lorenz}
\begin{figure}[h]
    \centering
       \includegraphics[width=0.3\textwidth]{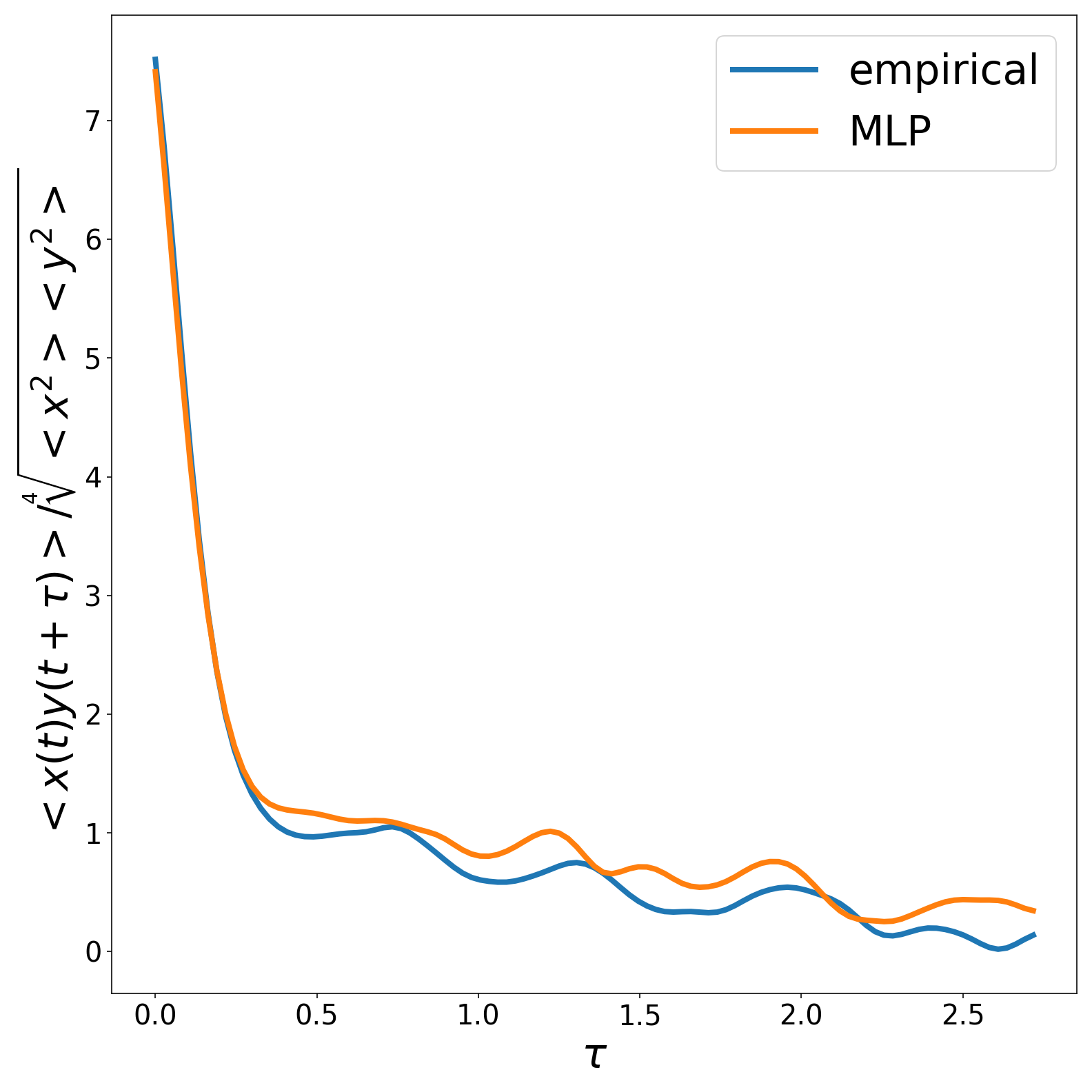}
        \includegraphics[width=0.3\textwidth]{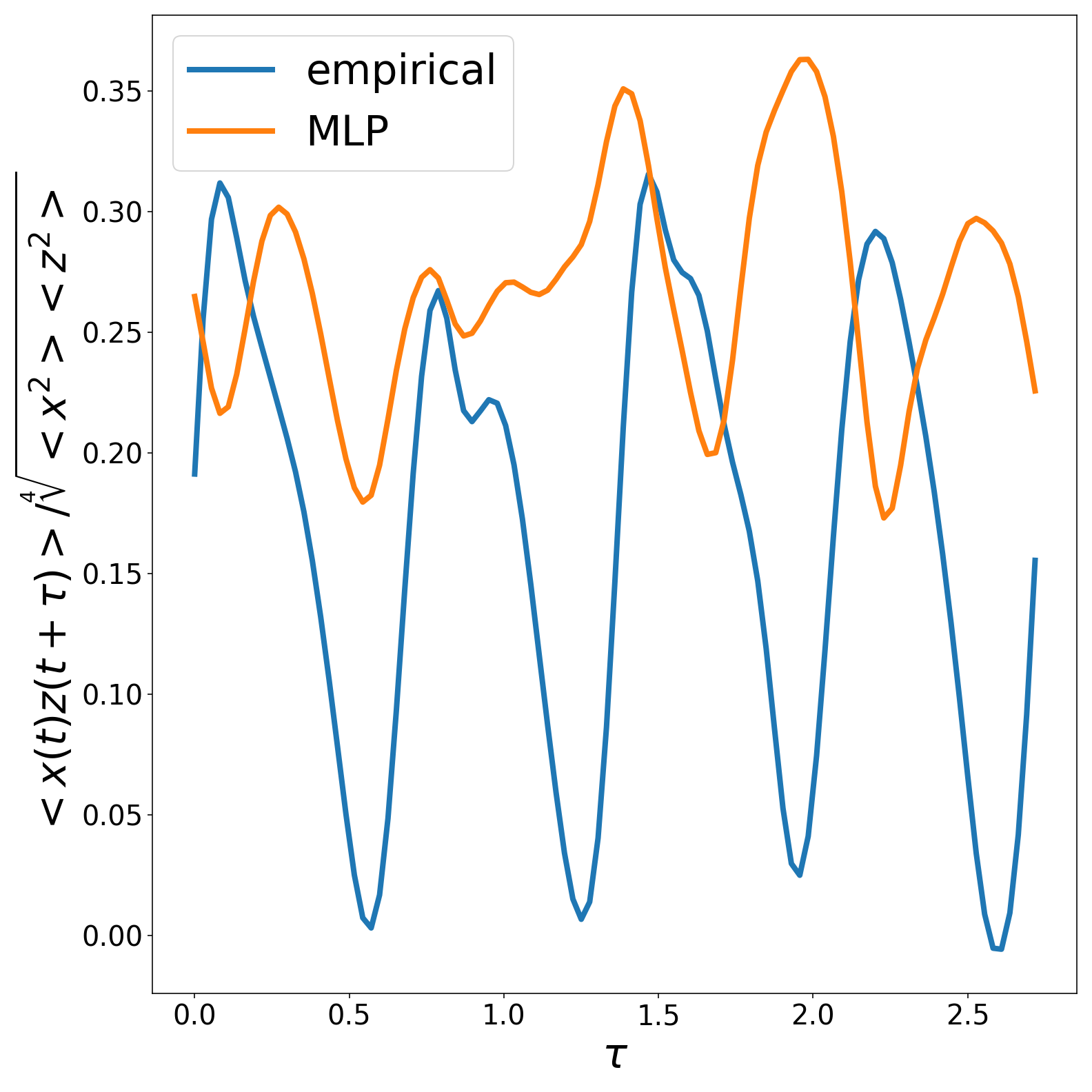}
      \includegraphics[width=0.3\textwidth]{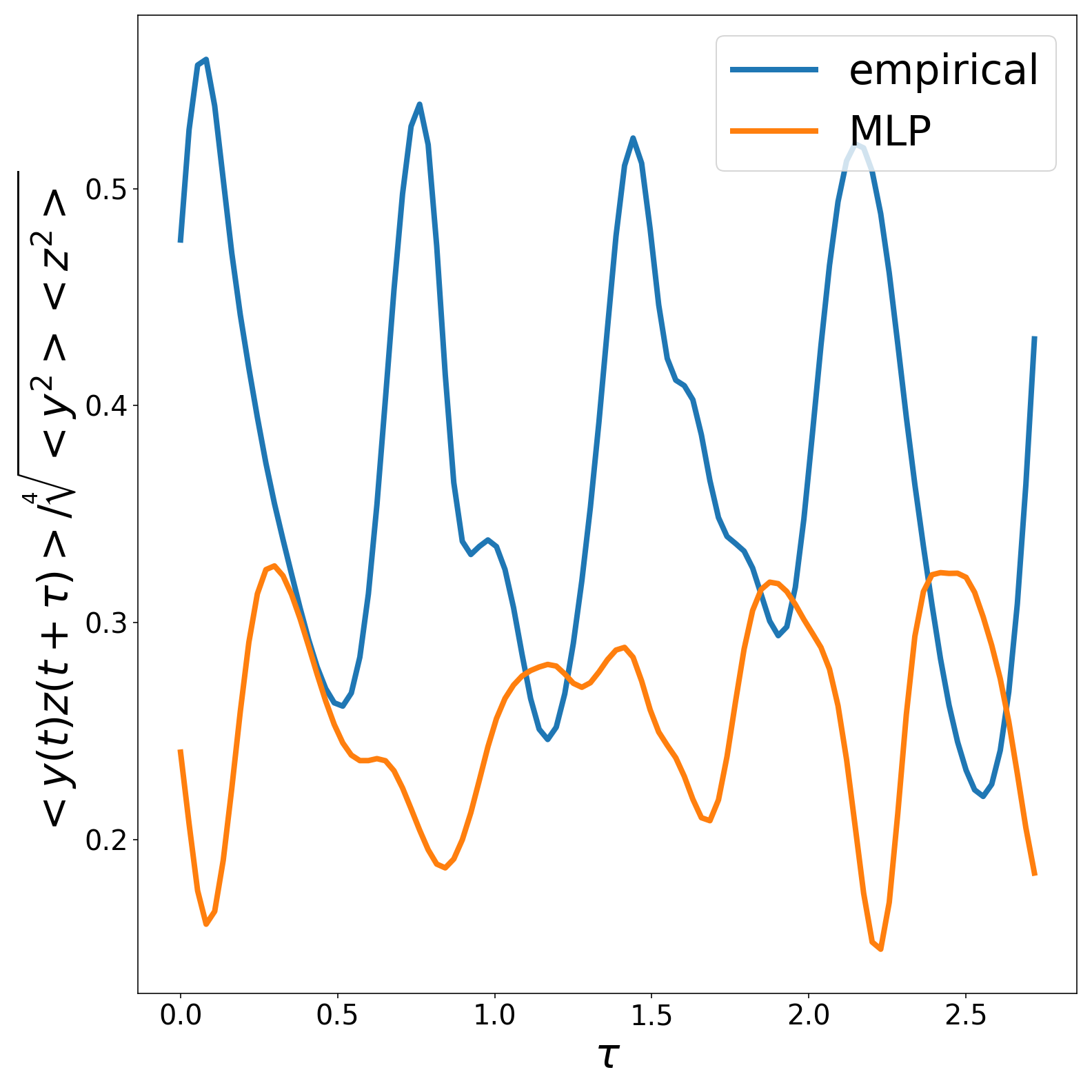}
      \includegraphics[width=0.3\textwidth]{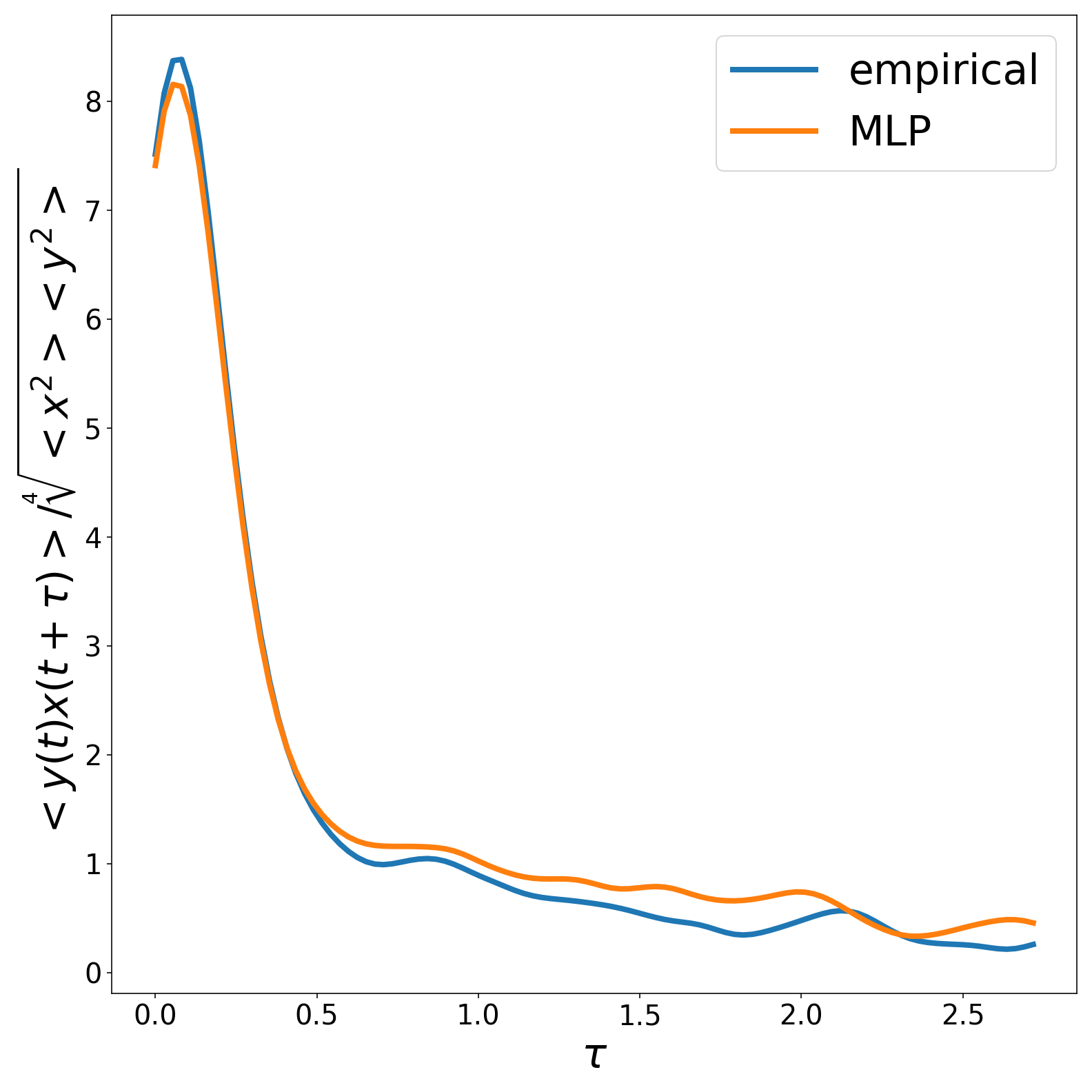}
       \includegraphics[width=0.3\textwidth]{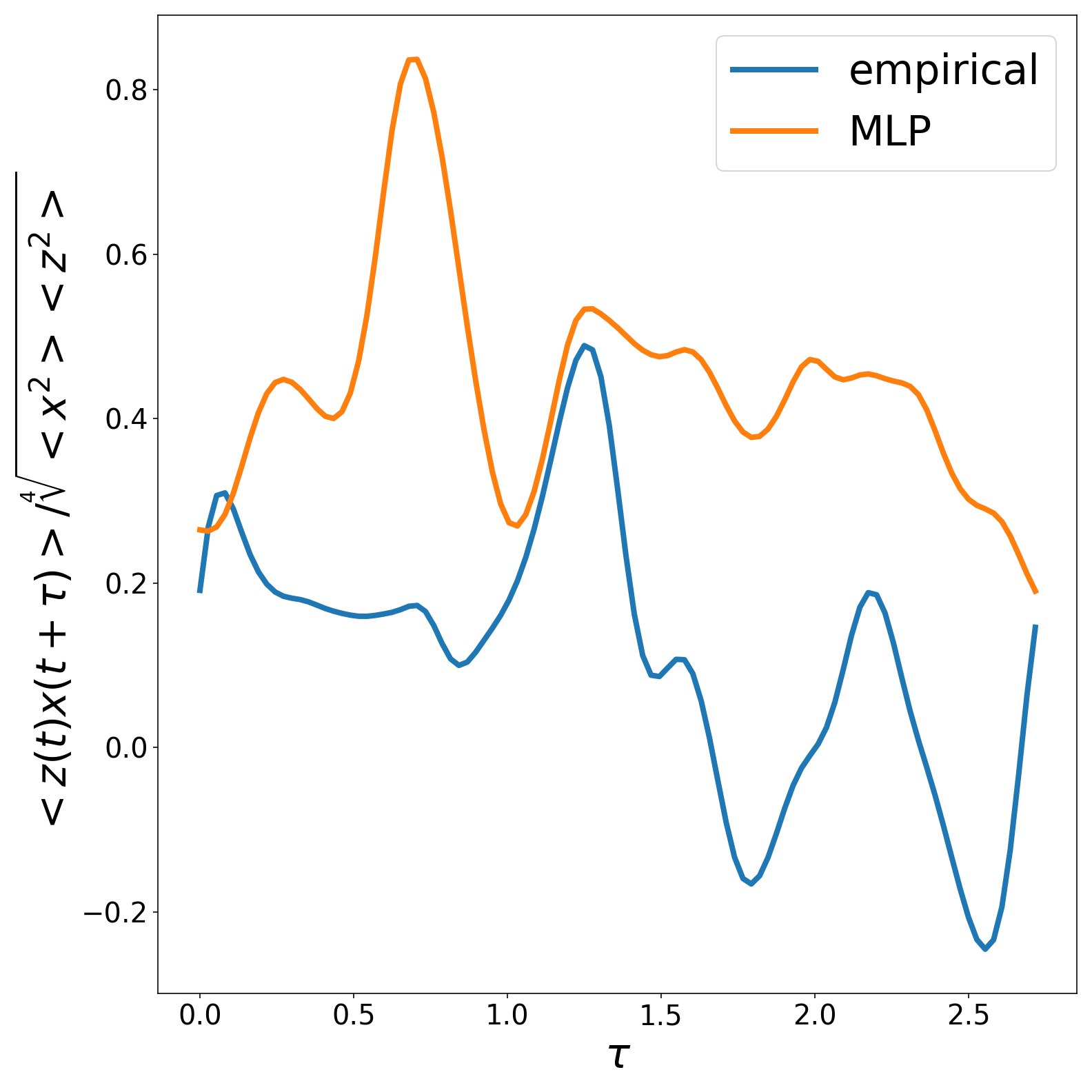}
       \includegraphics[width=0.3\textwidth]{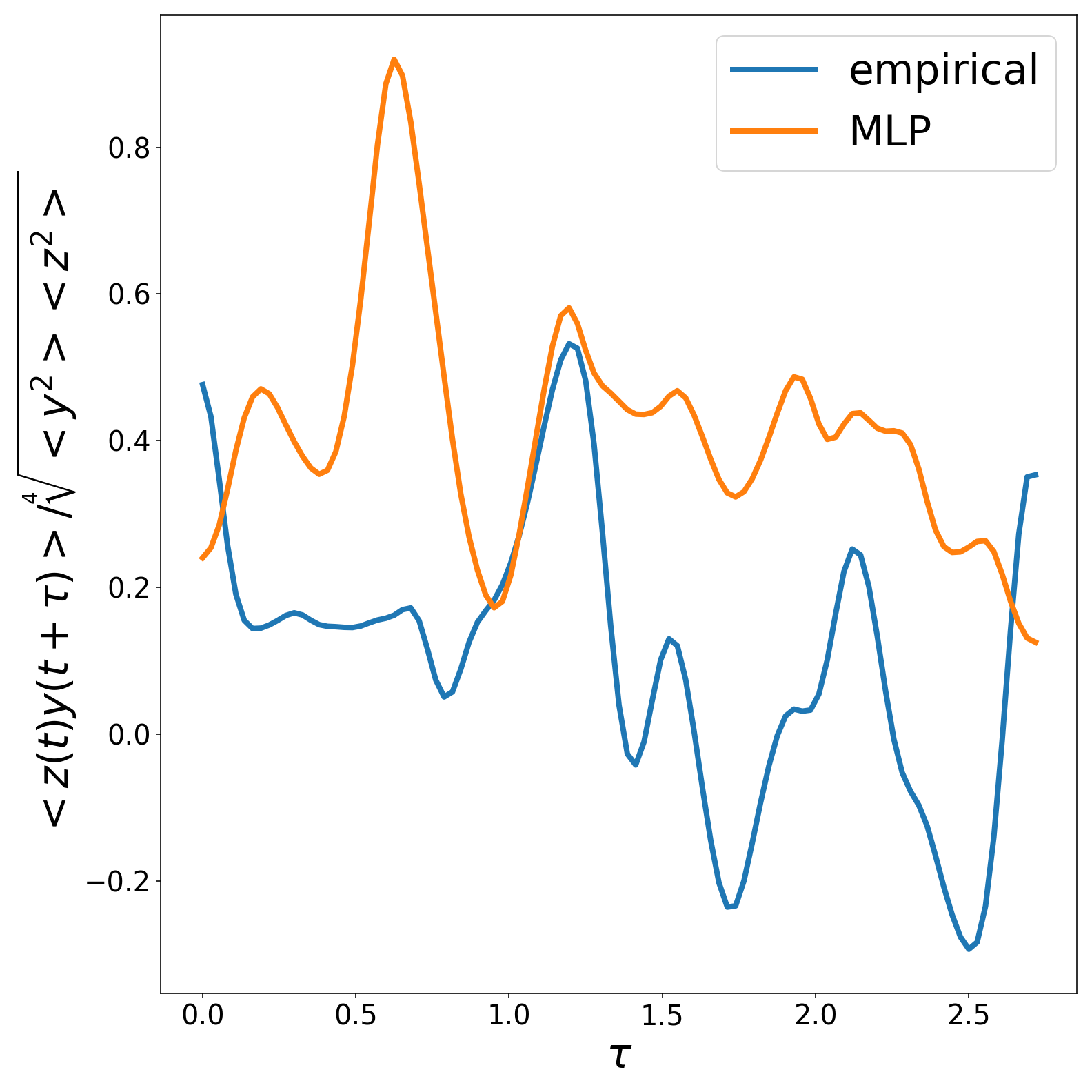}
    \caption{Normalized auto-correlation function $<x_i(t) x_j(t+\tau)>$ for the MLP model compared against that computed through direct integration of the Lorenz system Eq. (\ref{eq:eq-Lorenz}) (blue line). The time is expressed in terms of the Lyapunov time based on the maximum Lyapunov exponent. }
    \label{fig:app-crcorr}
\end{figure}
In this appendix, we provide some complementary results to those presented in section \ref{sec:chaotic_systems}.
In Fig. \ref{fig:app-crcorr}, we show the cross-correlations between the different variables on a long-time horizon.
We focus on the MLP architecture, since SINDy basically overlaps with the empirically simulated Lorenz system.
While the correlation between $x$ and $y$ are quite well reproduced, differences are displayed in the others where $z$ is present.
These correlations are indeed much weaker, and the ratio noise-signal larger.
In these cases, qualitative dynamics may be captured, but the difference is important even at short times. 
These results shows that the MLP is able to  well predict dominant contributions, which leads to an excellent reproduction of the phase-space dynamics. Yet, sub-dominant correlations are not well captured, and these errors eventually are magnified by chaotic dynamics leading to the difference in the Response function.

\section{Asymptotic regime of the causal response with random matrix theory}\label{App:RMT}
We are interested to obtain the asymptotic behavior of the LRR causal indicator in the so-called proportional scaling limit, namely when the number
of independent samples N goes to infinity, along with the number of features - i.e. the number of observed nodes $d$ in our case -  with fixed ratio
$\rho = N/d$. This can be obtained by some
random matrix formulas based on RMT~\cite{MP_law,ledoit2011eigenvectors}, which we simply state here.  
Let $\nu$ be the spectral distribution of $C$. The rescaled modulus $\vert\x_t\vert^2/d$  concentrates on a deterministic value in the large scale limit. 
This leads us to obtain a simpler form of the RMT equations than the general Marchenko-Pastur equations. 
Considering the quantity
\[
\Gamma \egaldef \lim_{N,P\to\infty} \frac{\alpha}{N}\Tr[G\n C],
\]
we get the following self-consistent equations:
\begin{align}
  \Gamma &= \frac{\alpha}{\rho}\int  \nu(dx)\frac{x}{1+\Lambda x} \label{eq:rmt_gamma}\\[0.2cm]
  \Lambda &= \frac{\alpha}{1+\Gamma}      \label{eq:rmt_lambda}
\end{align}
These in turn define a resolvent $G = \bigl(\I+\Lambda C)^{-1}$, where $\Lambda$ represents a self-energy when using the terminology of field theory.
$G$ represents a deterministic equivalent of $G\n$~\cite{hachem2007deterministic,chouard2023sample}, in the sense that for any sequence of $\R^d\times \R^d$ matrix $M$
of Frobenius norm equal to one, $\Tr\Bigl[(G-G\n)M\Bigr]$ goes to zero typically like $\sqrt{\log(N)/N}$. In order to leverage this we need first to get rid of factors of the form
${G\n}^2$ or $G\n\e_i\e_i^\top G\n$ present in equation~(\ref{eq:cr}).
For ${G\n}^2$ we make use of the following identity:
\[
{G\n}^2 = G\n+\alpha\frac{d}{d\alpha} G\n 
\]
This means that we need the derivative of $G$ with respect to $\alpha$:
\[
\frac{d}{d\alpha} G = \frac{\Lambda'(\alpha)}{\Lambda}(G^2-G)
\]
$\Lambda'(\alpha)$ can be obtained from the self-consistent equation~(\ref{eq:rmt_gamma},\ref{eq:rmt_lambda}). We have 
\[
\Lambda'(\alpha) = \frac{\Lambda^2}{\alpha^2}\frac{1}{1-Q}
\]
after introducing the quantity 
\[
Q = \frac{1}{\rho}\int \nu(dx) \frac{(\Lambda x)^2}{(1+\Lambda x)^2}.
\]
Note that $Q<1$ for $\rho>1$, while in the overparameterized regime ($\rho<1$) it can tend to $1$ when the $L_2$ regularization tends to zero, which results
in a divergence of the test error in that case (see e.g.~\cite{furtlehner2023free} for a thorough discussion and
interpretation of these quantities regarding overfitting).
For the second term, we use a similar trick by defining
\[
G_\gamma\n = \bigl(\I+\gamma\e_i\e_i^\top+\alpha C\n\bigr)^{-1},
\]
as a generating function. 
As can be shown the corresponding deterministic equivalent takes the form
\[
G_\gamma = \bigl(\I+\gamma\e_i\e_i^\top+ \Lambda_i(\gamma)C\bigr)^{-1},
\]
with $\Lambda_i$ solution of the fixed point equations~(\ref{eq:rmt_gamma},\ref{eq:rmt_lambda}) now depends on $\gamma$. Expressing
the fixed point equations formally at finite $N$ to make the expansion
with respect to $\gamma$ meaningful, we now have
\[
\Gamma(\gamma) = \frac{\alpha}{N}\Tr\Bigl[\frac{1}{\I+\gamma\e_i\e_i^\top+ \Lambda_i(\gamma)C} C\Bigr],
\]
while the factor  of interest now takes the desired form
\[
G\n\e_i\e_i^\top G\n  = \frac{d}{d\gamma}G_\gamma\n\Bigr\vert_{\gamma = 0},
\]
directly amenable to asymptotic analysis. Indeed the deterministic equivalent, which now depends on $\gamma$ can be inserted to yield 
\[
\frac{d}{d\gamma}G_\gamma\Bigr\vert_{\gamma = 0} = G\bigl(\e_i\e_i^\top+\Lambda_i'(0)C\bigr)G.
\]
Using the fixed point equation we get
\[
\Lambda_i'(0) = \frac{\Lambda}{N}\frac{\e_i^\top G(1-G)\e_i}{1-Q}.
\]
Assembling everything we finally get the following asymptotic estimation of the response in the proportional limit (up to leading ${\mathcal O}\bigl(1/N\bigr)$
contribution)
\begin{align}
  \chi_{i\to j}(t) &= \Tr\Bigl[(\I-G)\e_i\e_i^\top (\I-G){M^t}^\top\e_j\e_j^\top M^t\Bigr] \nonumber \\[0.2cm]
  &+ \frac{1}{N}\frac{1}{1-Q}\e_i^\top G(\I-G)\e_i\Bigl(\Tr\Bigl[G(\I-G){M^t}^\top \e_j\e_j^\top M^t\Bigr] 
  +\Lambda\sigma^2\sum_{\tau=0}^{t-1}
  \e_j^\top M^{t-\tau-1}{M^{t-\tau-1}}^\top\e_j \label{eq:cr_rmt}\Bigr)
\end{align}
The first two terms represent the deterministic (regularized) causal response  obtained in absence of noise. They basically corresponds to the (regularized) weighted sum of
round trips between $i$ and $j$. The last term is a variance term resulting from the noise.
Quantities like the overlap~(\ref{eq:overlap}) and $\Vert\RR_{\to i}\n(t)\Vert$
become deterministic asymptotically. We actually obtain
\begin{align}
\f_{\to i}^\top\RR_{\to i}(t) &= \e_i^\top M^t (\I-G){M^t}^\top\e_i,  \nonumber\\[0.2cm]
\Vert\RR_{\to i}\n(t)\Vert^2 &= \e_i^\top M^t(\I-G)^2{M^t}^\top\e_i + \Delta\Bigl[  \e_i^\top M^t G(\I-G){M^t}^\top \e_i 
  +\Lambda\sigma^2\sum_{\tau=0}^{t-1}
  \e_i^\top M^{t-\tau-1}{M^{t-\tau-1}}^\top\e_i \Bigr]\label{eq:overlap2}
\end{align}
with
\[
\Delta \egaldef \frac{1}{1-Q}\int\nu(dx) \frac{\Lambda x}{\bigl(1+\Lambda x\bigr)^2}.
\]

\end{document}